\documentclass[letterpaper,twocolumn,prr,
aps,showpacs,superscriptaddress,
floatfix]{revtex4-2}

\usepackage{tikz}
\usepackage{amssymb,amsmath}
\usepackage{graphicx}
\usepackage{booktabs}
\usepackage{verbatim}
\usepackage{lmodern} 
\usepackage{hyperref}
\usepackage{url}
\usepackage{csquotes}
\usepackage{epstopdf}
\usepackage{booktabs}
\usepackage{verbatim}
\usepackage[IL2]{fontenc}  
\usepackage[caption=false]{subfig}

\begin{document}

\begin{abstract}
Dynamical systems can display a plethora of ergodic and ergodicity breaking behaviors, ranging from simple periodicity to ergodicity and chaos. Here we report an unusual type of non-ergodic behavior in a many-body discrete-time dynamical system, specifically a multi-periodic response with multi-fractal distribution of equilibrium spectral weights at all rational frequencies. This phenomenon is observed in the momentum-conserving variant of the newly introduced class of the so-called parity check reversible cellular automata, which we define with respect to an arbitrary bi-partite lattice. Although the models display strong fragmentation of phase space of configurations, we demonstrate that the effect qualitatively persists within individual fragmented sectors, and even individual typical many-body trajectories. We provide detailed numerical analysis of examples on 2D (honeycomb, square) and 3D (cubic) lattices.
\end{abstract}

\title{Deterministic many-body dynamics with multifractal response}

\author{Yusuf Kasim}
\affiliation{Faculty of Mathematics and Physics, University of Ljubljana, Jadranska 19, SI-1000 Ljubljana, Slovenia}
\author{Toma\v{z} Prosen}
\affiliation{Faculty of Mathematics and Physics, University of Ljubljana, Jadranska 19, SI-1000 Ljubljana, Slovenia}
\affiliation{Institute of Mathematics, Physics and Mechanics, Jadranska 19, SI-1000 Ljubljana, Slovenia}
\date{\today}
\maketitle
 
\section{Introduction}
The ergodic hypothesis formulated by Boltzmann in the late 19th century, the main assertion of which is an equivalence between time and statistical ensemble averages, is one of the key principles laying the mathematical foundation of statistical mechanics developed by Birkhoff, Von Neumann and others~\cite{Emch2002Ch7}. Nevertheless, the richness of dynamical behavior of
generic deterministic or Hamiltonian systems to date could not be captured by
a unified and rigorous mathematical theory. In the 1950s, the famous Fermi-Pasta-Ulam-Tsingou model demonstrated that the ergodic hypothesis can not always hold~\cite{Fermi55} and paved the way to the development of the theory of solitons and integrable models. Furthermore, the Kolmogorov-Arnold-Moser (KAM) theory shows that generic Hamiltonian systems are neither integrable,  nor ergodic~\cite{Markus74}. One can classify dynamical systems on the basis of the interplay of determinism and randomness, building a mathematical structure of the so-called ergodic hierarchy~\cite{Berkovitz06,Sinai}. However, in order to have the physical picture, one should also consider and classify possible effects of ergodicity breaking, which is linked to physical phenomena such as symmetry breaking, phase transitions, and spin glasses~\cite{Venkataraman1989}. A system with broken ergodicity is confined to nontrivial subregions of phase space. 
This can be achieved by a structural mechanism such as free energy barriers or a dynamical mechanism by confining the time evolution by exact or approximate (cf. KAM tori) conserved quantities~\cite{Palmer82}. Understanding the physics of ergodicity breaking in classical many-body dynamics should have important implications to quantum many-body physics, where ergodic theory is much less developed. Nevertheless, it leads to important physical effects such as for instance: anomalous linear response and transport~\cite{Mazur,Zotos,Prosen,Bertini21}, 
quantum scars~\cite{Heller84,Kaplan_1999,Abanin}, fragmentation~\cite{Pollmann,Bernevig},
time-crystals~\cite{Wilczek12,Vedika,Nayak}, non-stationary dynamics and time-periodic equilibrium states~\cite{Buca}.

One of the more recent venues of studying many-body dynamics and non-equilibrium statistical mechanics is reversible cellular automata~\cite{Takesue,Bobenko}. They may support a plethora of different behavior between integrable and chaotic (ergodic) cellular automata, and have been used to provide exact solutions and analytical insights to transport and thermalization dynamics in interacting many-body systems~\cite{Buca_2021,Medenjak17,
Prosen_2016,
Klobas_2022,Klobas21,Wilkinson20,Pozsgay,Prosen2023,Rustem}. More recently, they have been crucial to the understanding of a universality class of anomalous fluctuations in integrable systems~\cite{Krajnik22,Krajnik24}.

Here we report the discovery of a class of discrete phase space deterministic reversible many-body systems --- parity check cellular automata on regular lattices (in two or more dimensions), with non-ergodic dynamics having multifractal dynamical structure factor and power-law distributions of AC Drude weights~\cite{Zadnik} at all rational frequencies. While cellular automata have been used before for constructing fractal or multifractal 
trajectories growing from specific initial conditions (non-equilibrium quench problem at zero entropy)~\cite{Nagler05,Murguia09,Hayase00,Claussen04}, we observe here multifractal equilibrium dynamics (at finite entropy density). From a large-scale hydrodynamic (or thermodynamic) perspective, our models describe a kind of multi-scale glass in fundamental distinction to seemingly similar HPP or HPF cellular automata~\cite{Hardy73,Hardy76,Frisch86} which model regular Euler hydrodynamics.

\section{Parity check automata}
Here we introduce a class of dynamical systems with discrete state space studied in this work.
Let $G=(V,E)$ be an undirected bipartite graph with 2-component set of
vertices $V=A\cup B$, $A\cap B=\emptyset$, such that each vertex $v\in A$ $(B)$ joins $n=n(v)$ vertices in $B$ ($A$) via edges $e(v)=(e_1,\ldots,e_n)$. We assign dynamical degree of freedom $s_e\in\mathbb Z_2$ to each of the $N=|E|$ edges, such that a configuration $\underline{s}\in\mathbb Z_2^N$
represents a unique state of cellular automaton, interpreting $\mathcal C=\mathbb Z_2^N$ as a space of all configurations, i.e. ``the phase space".
To each vertex, we assign a one-to-one map $\Phi_v : \mathbb Z^n_2 \to\mathbb Z^{n}_2$, 
for which at this point we only require to satisfy
the parity check (PC) property
\begin{equation}
(s'_1\dots s'_n)=\Phi_v(s_1\dots s_n) 
\; \Rightarrow \;
s_i+s_j=s_i'+s_j'\!\!\pmod{2},
\end{equation}
for all pairs $i,j\in\{1\ldots n\}$. We extend each local map $\Phi_v$ to $\mathcal C$ acting trivially (as identity) on all other components. As all maps 
$\{\Phi_v\}_{v\in C}$ mutually commute for fixed $C=A,B$, we define a reversible
cellular automaton (discrete dynamical system) with the full one-step update rule
\begin{equation}
\Phi = \prod_{v\in B}\Phi_v\prod_{v\in A}\Phi_v,\quad
\underline{s}(t+1) = \Phi (\underline{s}(t)).
\end{equation}
It may be useful to embed the set of vertices in $\mathbb R^d$
and think of an excitation on the edge/link $e$ as a unit mass particle carrying one unit of momentum $\vec{e}$ directed from $v$ to the neighboring vertex $v'$. In this respect, we shall study only regular lattices where all possible momenta from each fixed vertex add to zero. Motivated by possible physics applications, we will then be interested in momentum conserving PC automata (MCPCA) for which $(s'_1\dots s'_n)=\Phi_v(s_1\dots s_n)$ implies $\sum_{e=1}^n s_e \vec{e} = -\sum_{e=1}^n s'_e \vec{e}$.
For the simplest nontrivial bipartite lattice in $d=2$, i.e. honeycomb lattice with coordination number $n=3$ there is a unique nontrivial 
MCPCA defined by: $\Phi_v(s,s,s) = (s,s,s)$ and otherwise
$\Phi_v(s,s',s'') = (\bar{s},\bar{s}',\bar{s}'')$, where $\bar{s}=1-s$, illustrated graphically as~\footnote{There is exactly one more MCPCA rule with $n=3$, $\Phi_v(s,s,s)=(\bar{s},\bar{s},\bar{s})$, with trivial dynamics for which each configuration is exactly repeated after two time-steps.}
\begin{eqnarray}
	\begin{tikzpicture}[baseline={(current bounding box.center)},every node/.style={inner sep=0,outer sep=0},line cap=rect]
    \node (n0) at (0,0)[circle,draw,fill,inner sep=0.9375pt] {};
    \node (n1) at (0,0.75) {};
    \node (n2) at (-0.649519053,-0.375) {};
    \node (n3) at (0.649519053,-0.375) {};
    \draw[-] (n0) -- (n1);
    \draw[-] (n0) -- (n2);
    \draw[-] (n0) -- (n3);
    \node (A) at (0,0.375)[circle,draw,fill=white,inner sep=2.625pt]{};
    \node (C) at (-0.324759526,-0.1875)[circle,draw,fill=white,inner sep=2.625pt]{};
    \node (D) at (0.324759526,-0.1875)[circle,,draw,fill=white,inner sep=2.625pt]{};
    \end{tikzpicture}
	\leftrightarrow
	\begin{tikzpicture}[baseline={(current bounding box.center)},every node/.style={inner sep=0,outer sep=0},line cap=rect]
    \node (n0) at (0,0)[circle,draw,fill,inner sep=0.9375pt] {};
    \node (n1) at (0,0.75) {};
    \node (n2) at (-0.649519053,-0.375) {};
    \node (n3) at (0.649519053,-0.375) {};
    \draw[-] (n0) -- (n1);
    \draw[-] (n0) -- (n2);
    \draw[-] (n0) -- (n3);
    \node (A) at (0,0.375)[circle,draw,fill=white,inner sep=2.625pt]{};
    \node (C) at (-0.324759526,-0.1875)[circle,draw,fill=white,inner sep=2.625pt]{};
    \node (D) at (0.324759526,-0.1875)[circle,,draw,fill=white,inner sep=2.625pt]{};
    \end{tikzpicture}
    &\qquad&
	\begin{tikzpicture}[baseline={(current bounding box.center)},every node/.style={inner sep=0,outer sep=0},line cap=rect]
    \node (n0) at (0,0)[circle,draw,fill,inner sep=0.9375pt] {};
    \node (n1) at (0,0.75) {};
    \node (n2) at (-0.649519053,-0.375) {};
    \node (n3) at (0.649519053,-0.375) {};
    \draw[-] (n0) -- (n1);
    \draw[-] (n0) -- (n2);
    \draw[-] (n0) -- (n3);
    \node (A) at (0,0.375)[circle,draw,fill=red,inner sep=2.625pt]{};
    \node (C) at (-0.324759526,-0.1875)[circle,draw,fill=red,inner sep=2.625pt]{};
    \node (D) at (0.324759526,-0.1875)[circle,,draw,fill=red,inner sep=2.625pt]{};
    \end{tikzpicture}
	\leftrightarrow
    \begin{tikzpicture}[baseline={(current bounding box.center)},every node/.style={inner sep=0,outer sep=0},line cap=rect]
    \node (n0) at (0,0)[circle,draw,fill,inner sep=0.9375pt] {};
    \node (n1) at (0,0.75) {};
    \node (n2) at (-0.649519053,-0.375) {};
    \node (n3) at (0.649519053,-0.375) {};
    \draw[-] (n0) -- (n1);
    \draw[-] (n0) -- (n2);
    \draw[-] (n0) -- (n3);
    \node (A) at (0,0.375)[circle,draw,fill=red,inner sep=2.625pt]{};
    \node (C) at (-0.324759526,-0.1875)[circle,draw,fill=red,inner sep=2.625pt]{};
    \node (D) at (0.324759526,-0.1875)[circle,,draw,fill=red,inner sep=2.625pt]{};
    \end{tikzpicture}
    \nonumber \\
    	\begin{tikzpicture}[baseline={(current bounding box.center)},every node/.style={inner sep=0,outer sep=0},line cap=rect]
    \node (n0) at (0,0)[circle,draw,fill,inner sep=0.9375pt] {};
    \node (n1) at (0,0.75) {};
    \node (n2) at (-0.649519053,-0.375) {};
    \node (n3) at (0.649519053,-0.375) {};
    \draw[-] (n0) -- (n1);
    \draw[-] (n0) -- (n2);
    \draw[-] (n0) -- (n3);
    \node (A) at (0,0.375)[circle,draw,fill=red,inner sep=2.625pt]{};
    \node (C) at (-0.324759526,-0.1875)[circle,draw,fill=white,inner sep=2.625pt]{};
    \node (D) at (0.324759526,-0.1875)[circle,,draw,fill=white,inner sep=2.625pt]{};
    \end{tikzpicture}
	\leftrightarrow
	\begin{tikzpicture}[baseline={(current bounding box.center)},every node/.style={inner sep=0,outer sep=0},line cap=rect]
    \node (n0) at (0,0)[circle,draw,fill,inner sep=0.9375pt] {};
    \node (n1) at (0,0.75) {};
    \node (n2) at (-0.649519053,-0.375) {};
    \node (n3) at (0.649519053,-0.375) {};
    \draw[-] (n0) -- (n1);
    \draw[-] (n0) -- (n2);
    \draw[-] (n0) -- (n3);
    \node (A) at (0,0.375)[circle,draw,fill=white,inner sep=2.625pt]{};
    \node (C) at (-0.324759526,-0.1875)[circle,draw,fill=red,inner sep=2.625pt]{};
    \node (D) at (0.324759526,-0.1875)[circle,,draw,fill=red,inner sep=2.625pt]{};
    \end{tikzpicture}
    &\qquad& \textrm{$+\;\; 2\pi/3$ rotations} 
    \label{eq:rules}
\end{eqnarray}
While classifying general PC and MCPC reversible cellular automata will be the subject of a separate study, we shall here only extend the above rule to arbitrary lattice or graph geometry with vertices of coordination number $n$ as $\Phi_v(s,s\ldots s) = (s,s\ldots s)$ and otherwise
$\Phi_v(s_1,s_2\ldots s_n) = (\bar{s}_1,\bar{s}_2\ldots \bar{s}_n)$,
specifically for square lattice with $n=4$ and cubic lattice ($d=3$) with $n=6$. Our MCPCA has several key properties: (i) It conserves $d$ component total momentum vector $\vec{P}(\underline{s})=\sum_{e=1}^N
s_e \vec{e}$, $\vec{P} = \vec{P}\circ \Phi$. (ii) The parity of a sum of dynamical variables along any closed loop $\mathcal{L}=(e_1,e_2\ldots e_\ell)$, $\pi_\mathcal{L}(\underline{s}) = (-1)^{\sum_{i=1}^\ell s_{e_i}}$, is conserved $\pi_\mathcal{L} \equiv \pi_\mathcal{L} \circ \Phi$. (iii) If the classical spins along a closed loop $\mathcal{L}$ are in a Néel configuration 
$(s_{e_1},s_{e_2}\ldots s_{e_\ell})=(0,1,0,1\ldots 0,1)$ ($\ell$ even) then these spins are preserved at all times, i.e. the Néel loop is frozen~\cite{PavelOrlov}.
Specifically, the Néel loop is flipped after each half time-step.
All these properties follow from the definitions of the MCPCA. The property (ii) is particularly important: suppose a lattice 
with $N$ edges has $M$ independent closed loops, say along the smallest plaquettes like in the honeycomb lattice:
\begin{eqnarray}
\begin{tikzpicture}[baseline={(current bounding box.center)},every node/.style={inner sep=0,outer sep=0},line cap=rect,scale=0.7]
	\node (n0) at (0,0)[circle,fill,inner sep=1.25pt] {};
    \node (n1) at (0,0.5) {};
    \node (n2) at (-0.8660254037844387,-0.5)[circle,draw,inner sep=1.25pt] {};
	\node (n3) at (0.8660254037844387,-0.5)[circle,draw,inner sep=1.25pt] {};
    \node (n4) at (0.8660254037844387,-1.5)[circle,fill,inner sep=1.25pt] {};
	\node (n5) at (-0.8660254037844387,-1.5)[circle,fill,inner sep=1.25pt] {};
	\node (n6) at (0,-2)[circle,draw,inner sep=1.25pt] {};
	\node (n7) at (1.299038105676658,-0.25) {};
	\node (n8) at (-1.299038105676658,-0.25) {};
	\node (n9) at (1.299038105676658,-1.75) {};
	\node (n10) at (-1.299038105676658,-1.75) {};
	\node (n11) at (0,-2.5) {};
    \draw[-] (n0) -- (n1);
    \draw[-] (n0) -- (n2);
    \draw[-] (n0) -- (n3);
    \draw[-] (n3) -- (n4);
    \draw[-] (n2) -- (n5);
    \draw[-] (n5) -- (n6);
    \draw[-] (n4) -- (n6);
    \draw[-] (n6) -- (n11);
    \draw[-] (n4) -- (n9);
    \draw[-] (n5) -- (n10);
    \draw[-] (n3) -- (n7);
    \draw[-] (n2) -- (n8);
    \node (ar) at (0,-1) {$\circlearrowleft$};
	\draw[black,fill=red] (-0.43301270189221935cm+3.535533906pt,-0.25cm+3.535533906pt) arc  (45:225:5pt); 
	\draw[black,fill=white] (-0.43301270189221935cm-3.535533906pt,-0.25cm-3.535533906pt) arc  (225:405:5pt);
	\draw[black,fill=red] (0.43301270189221935cm+3.535533906pt,-0.25cm+3.535533906pt) arc  (45:225:5pt); 
	\draw[black,fill=white] (0.43301270189221935cm-3.535533906pt,-0.25cm-3.535533906pt) arc  (225:405:5pt);
	\draw[black,fill=red] (-0.8660254037844387cm+3.535533906pt,-1cm+3.535533906pt) arc  (45:225:5pt); 
	\draw[black,fill=white] (-0.8660254037844387cm-3.535533906pt,-1cm-3.535533906pt) arc  (225:405:5pt);
	\draw[black,fill=red] (0.8660254037844387cm+3.535533906pt,-1cm+3.535533906pt) arc  (45:225:5pt); 
	\draw[black,fill=white] (0.8660254037844387cm-3.535533906pt,-1cm-3.535533906pt) arc  (225:405:5pt);
	\draw[black,fill=red] (-0.43301270189221935cm+3.535533906pt,-1.75cm+3.535533906pt) arc  (45:225:5pt); 
	\draw[black,fill=white] (-0.43301270189221935cm-3.535533906pt,-1.75cm-3.535533906pt) arc  (225:405:5pt);
	\draw[black,fill=red] (0.43301270189221935cm+3.535533906pt,-1.75cm+3.535533906pt) arc  (45:225:5pt); 
	\draw[black,fill=white] (0.43301270189221935cm-3.535533906pt,-1.75cm-3.535533906pt) arc  (225:405:5pt);
	\end{tikzpicture}
	\label{eq:parity}
\end{eqnarray}
For honeycomb, square and cubic lattices, $M=N/3$, $M=N/2$, and $M=5N/6$,
respectively, so we expect that $\mathcal C$ fragments into $2^M$
sectors with fixed plaquette parities, leaving the components which still contain exponentially many, $2^{N-M}$, configurations.
We note that the properties (i) and (iii) do not significantly or qualitatively further reduce the size of fragmented components, which in any case remain exponential in $N$. For instance, we verified that the length of typical and longest periodic orbits are exponential in $N$.

\section{All-periodic multifractal dynamical response}
Fundamental observables characterizing ergodic properties of a
reversible cellular automaton are dynamical correlation functions of a local observable $\rho_v({\underline s})$
\begin{equation}
C_{\vec r}(t) = \langle 
\rho_v(\underline{s}) \rho_{v+\vec{r}}(\Phi^{(t)}(\underline{s})) \rangle,
\label{eq:dendencorrelation}
\end{equation}
where $\langle A\rangle\equiv\sum_{\underline s\in\mathcal C} p(\underline{s}) A(\underline{s})$ and $p$ represents an invariant (equilibrium) statistical state $p=p\circ \Phi$, $p(\underline{s})\ge 0$, $\sum_{\underline{s}} p(\underline{s})=1$. Throughout this work, we will consider a maximum entropy (`infinite temperature') ensemble with $p(\underline{s}) = 1/|\mathcal C|=2^{-N}$ and symmetrized local particle density
$\rho_v(\underline{s}) = \sum_{i=1}^{n(v)} s_{e_i(v)} - n(v)/2$
satisfying $\langle \rho_v \rangle = 0$. Here $\vec{r}$ denotes a valid displacement between a vertex $v$ and another vertex in the Euclidean embedding. The simplest and most informative is the local autocorrelator $\vec{r}=0$ which we will characterize in terms of a spectral function
\begin{equation}
	S(\omega) =  \sum_{t\in \mathbb{Z}} C_{\vec{0}}(t) \exp(2\pi i t \omega).
\end{equation}
Allowing for the possibility that the system has a sub-harmonic response at any integer periodicity we can write the spectral function as:
\begin{equation}
	S(\omega) = \sum'_{n,p} A_{n,p} \delta\Bigl(\omega - \frac{n}{p}\Bigr) + S_{\mathrm{reg}}(\omega),
\label{eq:Spectra_NN}
\end{equation}
where the sub-harmonic AC `Drude weights' read
\begin{equation}
	A_{n,p} = \lim_{T\to \infty} \frac{1}{T} \sum_{t=0}^{T-1} C_{\vec{0}}(t) \exp\Bigl(-\frac{2\pi i t n}{p}\Bigr).
 \label{eq:Aqp}
\end{equation}
The sum $\sum'_{n,p}$ runs over $n,p\in\mathbb Z$, $|n|<p$,
with $n,p$ coprime. In general, for ergodic and mixing dynamics, all $A_{n,p}\equiv 0$. For non-ergodic dynamics, $A_{0,1}\neq 0$, 
while for $p-$periodic dynamics (aka time crystal with $p$-subharmonic response~\cite{Nayak}), $A_{n,p}\neq 0$ for some $n$. 

We define dynamics to be {\em infinite-periodic} if for some (local) observable infinitely many $A_{n,p}\neq 0$, and {\em all-periodic} if $A_{n,p}\neq 0$ for all coprime $n,p$. As we shall demonstrate below, MCPCA represents a wide class of all-periodic dynamics, in the thermodynamic limit $N\to \infty$, with 
multifractal spectral distributions $S(\omega)$.

 \begin{figure}
	\centering
	\includegraphics[width=1.02\linewidth]{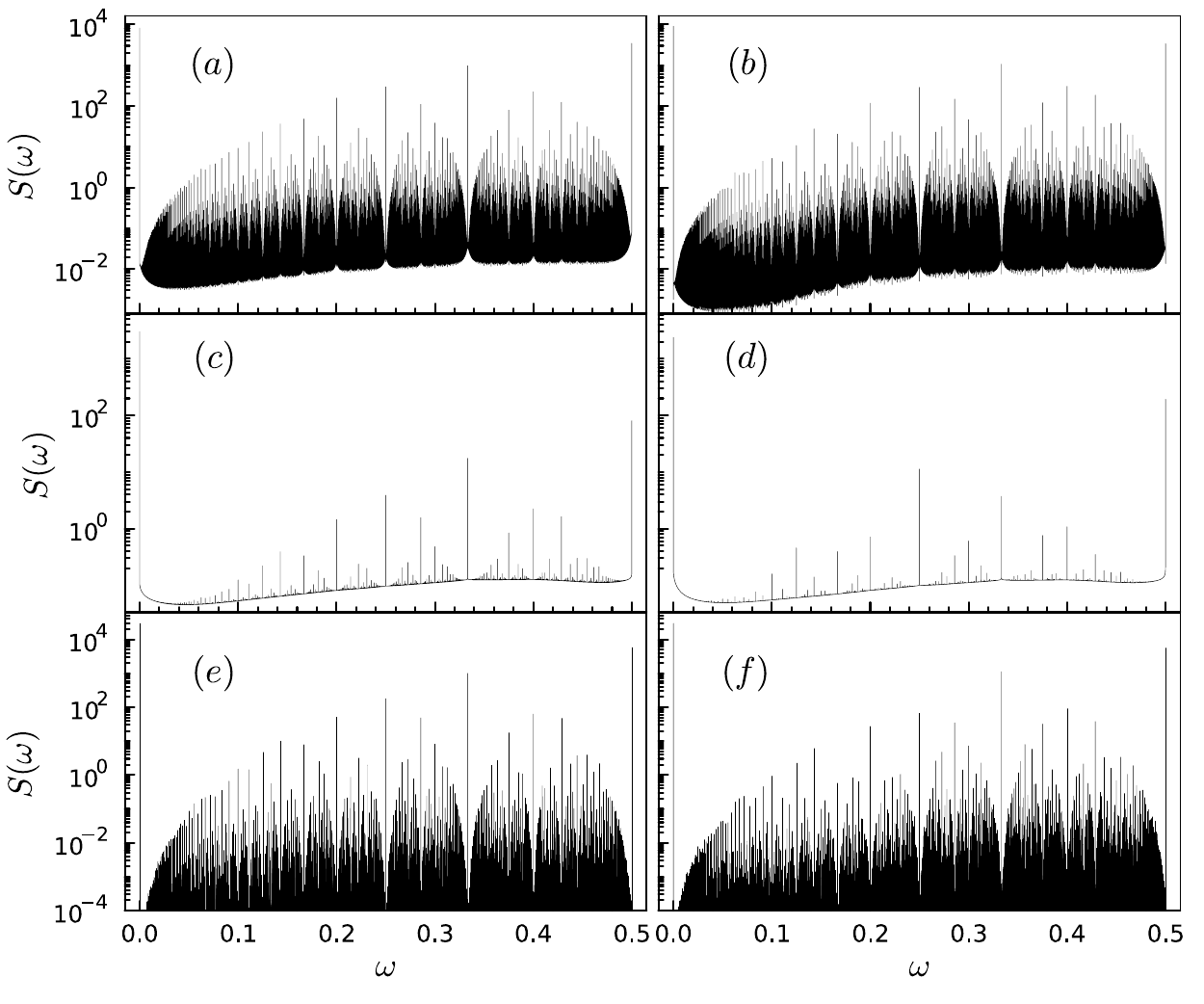}
	\caption{The power spectra of the density-density correlation function $S(\omega)$. All the plots are calculated for $T=2^{20}$ and using a Gaussian filter (see text for details).
 We show data for $16\times 16$ square lattice, $N=2\times 16^2$, in the random loop-parity sector (a) averaging over $\mathcal{N}=0.68\cdot 10^6$  random initial conditions, and in all-positive loop-parity sector (b) with $\mathcal{N}=0.80\cdot 10^6$;
for $16\times 16$-vertex honeycomb lattice, $N=(3/2)\times 16^2$,
within random (c), and all-positive (d) loop-parity sector, both with $\mathcal{N}=2.8\cdot 10^6$;
and finally, for $8\times 8\times 8$ cubic lattice, $N=3\times 8^3$,
within random (e) ($\mathcal N=1.2\cdot 10^6$), and all-positive  
(f) ($\mathcal N=0.65\cdot 10^6$) loop parity sector.}
	\label{fig:spectrum_full}
\end{figure}

\begin{figure}
	\centering
	\includegraphics[width=0.9\linewidth]{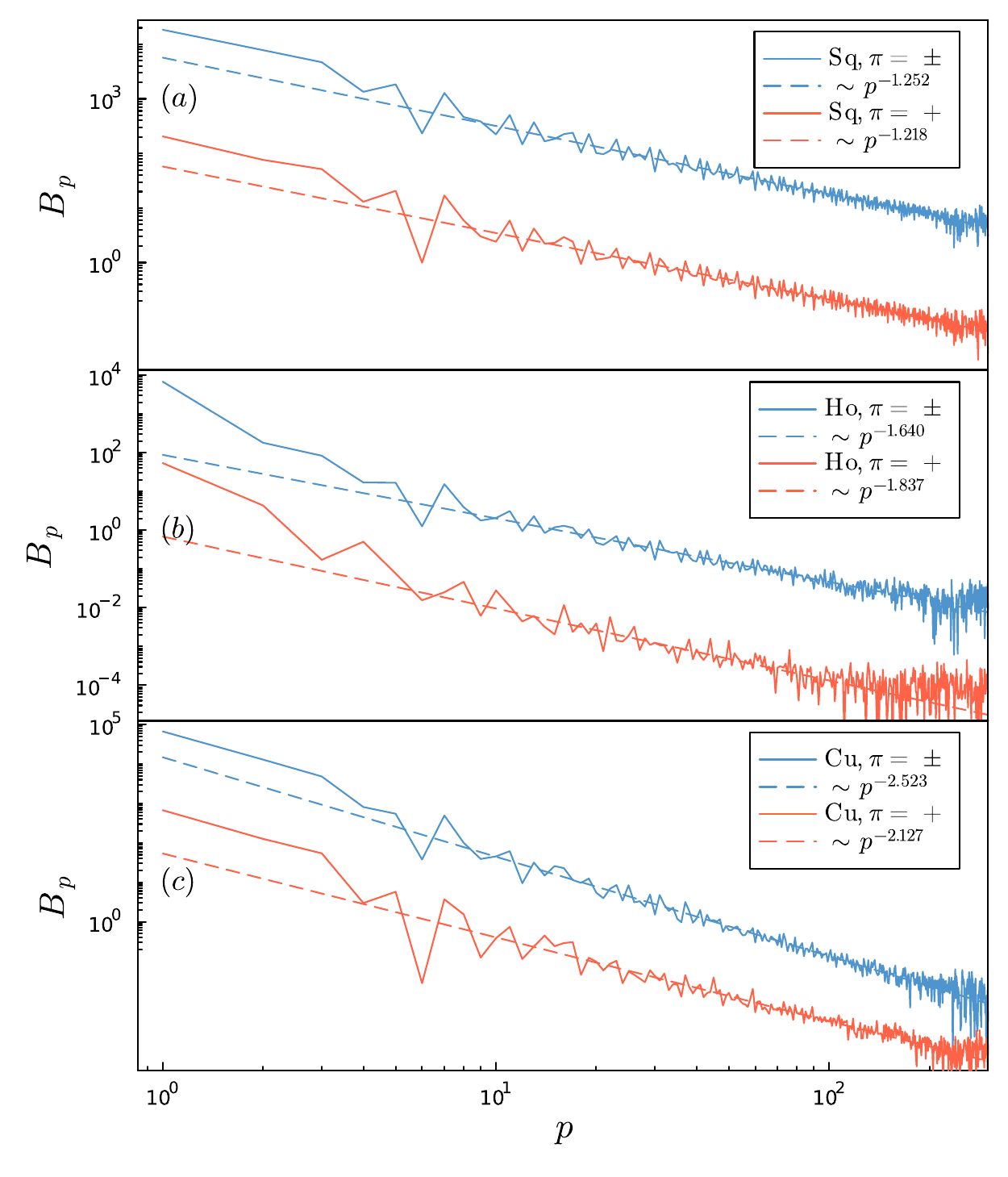}
	\caption{Total period-$p$ spectral weight $B_p= \sum_n A_{n,p}$ as a function of $p$ for (a) square, (b) honeycomb, and (c) cubic lattices for random loop-parity ($\pi=\pm$, in blue) and for positive parity($\pi=+$, in red) with best power law fits plotted in dashed (see legend). Note that the data for the positive loop-parity sector is shifted by factor $10^{-2}$ for better visualization. We used the data shown in Fig.~\ref{fig:spectrum_full} to produce these plots.}
	\label{fig:B_q}
\end{figure}

\begin{figure}
	\centering
	\includegraphics[width=0.9\linewidth]{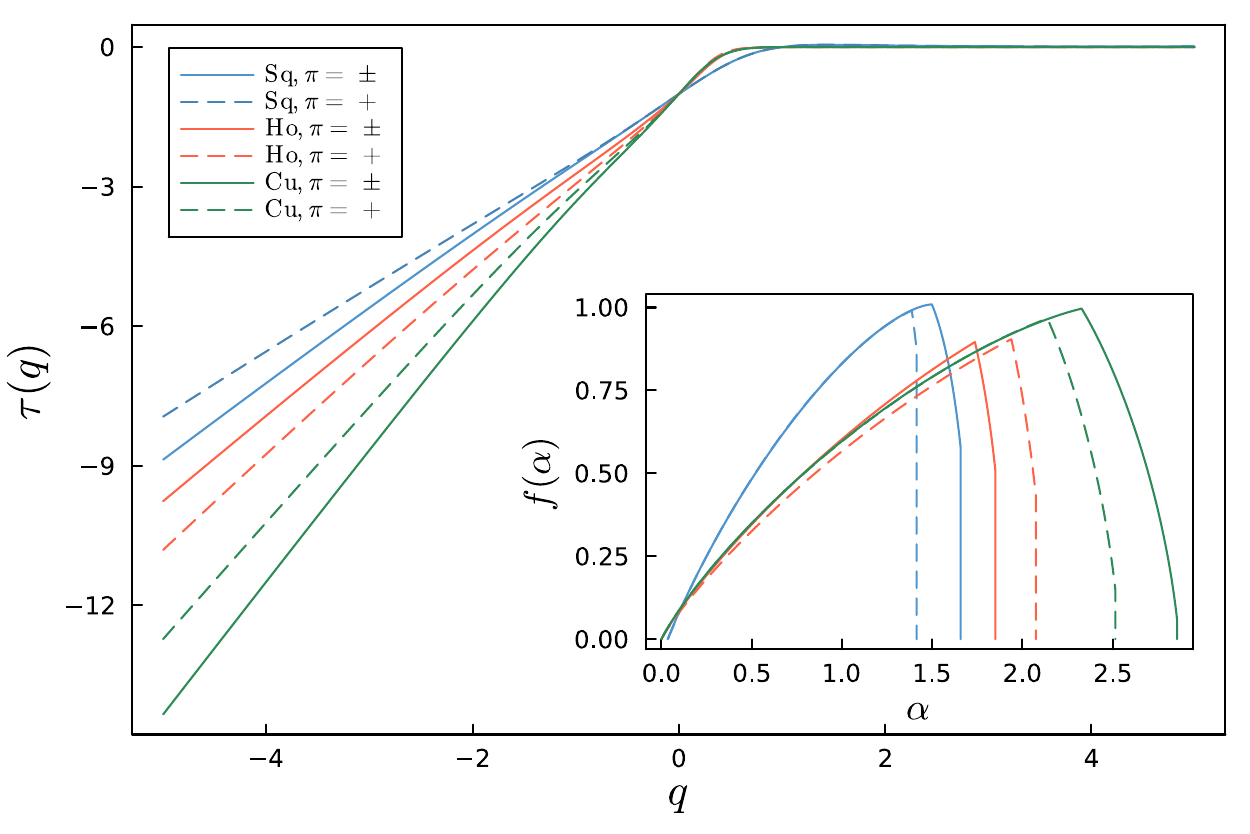}
	\caption{Moment scaling exponent $\tau (q)$ for square (Sq), honeycomb  (Ho), and cubic (Cu) lattices for random loop-parity ($\pi=\pm$, full lines) and all-positive loop-parity ($\pi=+$, dashed lines) sectors, for data shown in Fig.~\ref{fig:spectrum_full}. The inset shows the singularity spectrum  $f(\alpha)$. 
    }
	\label{fig:mutlifrac}
\end{figure}

\begin{figure}
	\centering
	\includegraphics[width=0.47\linewidth]{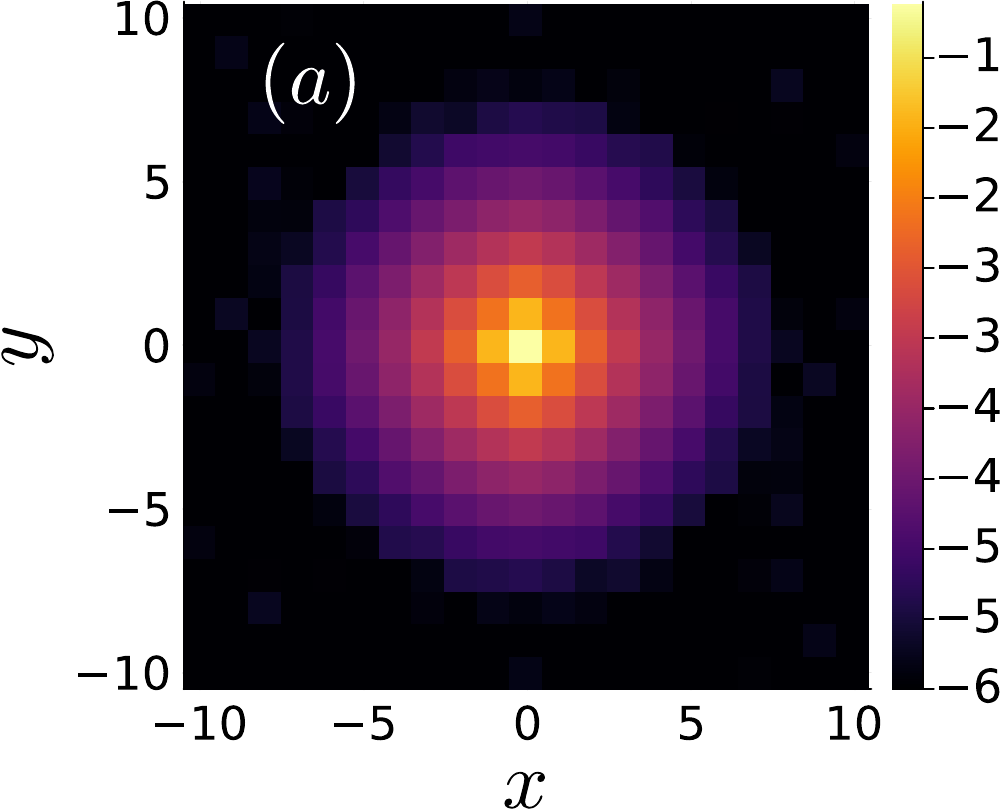}
    \vspace{-1mm}\hspace{1mm}\includegraphics[width=0.44\linewidth,height=0.38\linewidth]{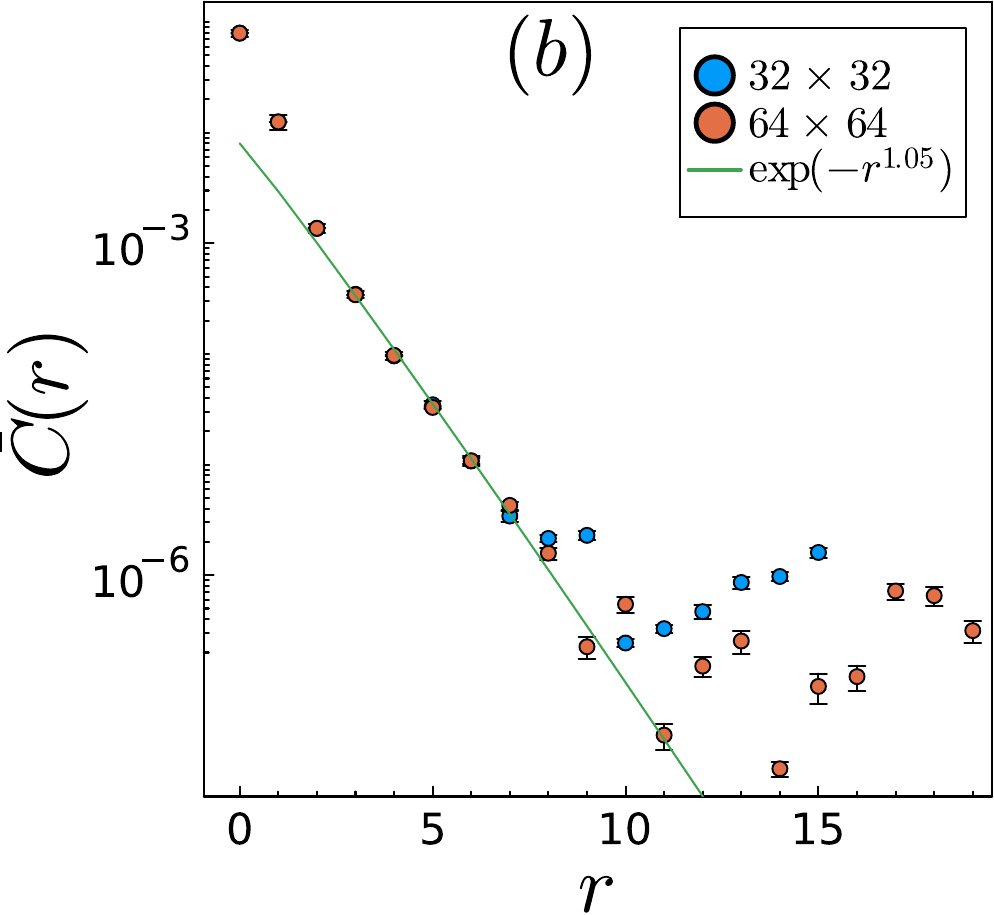}
	\caption{(a) Heatmap of the time averaged density-density correlation function 
    $\bar{C}(\vec{r})$ in log-10 scale for MCPCA on $64 \times 64$ square lattice. (b) 
    Radial behavior of $\bar{C}(\vec{r})$ averaged over four directions $(r,0),(-r,0),(0,r),(0,-r)$, and comparing two different lattice sizes, $32\times32$ (blue), and $64\times 64$ (red) and best fitted exponential decay $e^{-1.05 r}$ (green).
    For both lattice sizes, we averaged over $T=2048$ consecutive time-steps
    and sampled over $\mathcal{N} \sim 10^6$ random initial conditions in the random parity sector $\pi=\pm$ (estimated statistical error bars indicated).}
	\label{fig:TACF}
\end{figure}

\section{Results} 
We ran extensive Monte-Carlo simulations of the spectral function $S(\omega)$ for different regular bipartite lattices in $d=2,3$, honeycomb, square and cubic. We sampled initial configurations either fully randomly, i.e. considered flat equilibrium state $p(\underline{s})=2^{-N}$, or we considered a subsector of equally weighted $2^M$ configurations with all loop parities $\pi_{\mathcal L}=+1$.
These data sets are respectively labeled by $\pi=\pm$ and $\pi=+$.
In order to be able to resolve possible delta singularities in $S(\omega)$ more reliably, we use a Gaussian filtering of very long signals, $T=2^{20}$, and compute power spectrum as $S(\nu/T)=\sum_{t=0}^{T-1}
C_{\vec{0}}(t)e^{-16(t/T-1/2)^2+2\pi i t \nu/T}$, $\nu=-\frac{T}{2},-\frac{T}{2}+1\ldots\frac{T}{2}-1$,
using Fast Fourier Transform and average over ${\cal N}$ (typically ${\cal N}\approx 10^6$) initial configurations. The results are shown in Fig.~\ref{fig:spectrum_full}
and have been checked to be practically insensitive to system size once the number of edges $N$ exceeded a few hundred or so. 
Remarkably, the result is also robust if we replace the ensemble average by a time average of $S(\omega)$ computed over long sections of even longer fixed, typical trajectory (for sufficiently large $N$), as shown in App.~\ref{apx:Time_average}.
This means that $S(\omega)$, despite being a highly singular and irregular object, behaves like an ergodic observable in a fixed loop-parity phase space sector.

We observe a clear evidence of delta-spikes at all rational frequencies in all $S(\omega)$. Due to Gaussian filtering it was possible to reliably determine the weights $A_{n,p}$ for periods up to $p\approx 300$ by polynomial $\varepsilon\to 0$ extrapolations of
$\int_{n/p-\varepsilon}^{n/p+\varepsilon}{\rm d}\omega S(\omega) = A_{n,p} + A'_{n,p}\varepsilon+\ldots$  Writing the total weight of period $p$ as $B_p = \sum_n A_{n,p}$
we find a very clear evidence of asymptotic power law scaling
\begin{equation}
B_p \asymp p^{-\mu},
\end{equation}
where the exponent $\mu$ depends only on the lattice geometry and phase space sector (see Fig.~\ref{fig:B_q}). We note that in all cases $\mu>1$, so the total singular contribution to $S(\omega)$ converges. We also find 
a significant nonzero regular part $S_{\rm reg}(\omega)$, except in the cubic lattice case, where it could possibly vanish. 

We also perform a multifractal scaling analysis of the power spectrum, specifically the singular part $S_{\rm sing}(\omega)=S(\omega)-S_{\rm reg}(\omega)$ which we compute as described in the previous paragraph. Splitting the interval $\omega \in [0,1]$ into $1/\varepsilon$ boxes of size $\varepsilon$, we denote the weight in the $i$-th interval as $P_i(\varepsilon)=\int_{i \varepsilon}^{(i+1)\varepsilon} {\rm d}\omega S_{\rm sing}(\omega)$. We then define the standard $\tau$-exponents, dependent on $q\in \mathbb R$, as:
\begin{equation}
\label{eq:tau_q}
\tau(q) = \lim_{\varepsilon\to 0}\frac{\log\sum_i [P_i(\varepsilon)]^q}{\log\varepsilon}
\end{equation}
where in practice we make sure that $\varepsilon$ is much bigger than frequency resolution $1/T \ll \varepsilon \ll 1$.
Finally, multifractal spectrum is determined as $f(\alpha) = {\rm min}_q (q \alpha - \tau(q))$, see Fig.~\ref{fig:mutlifrac}, and App.~\ref{apx:details_tau} for details on calculation of $\tau(q)$ and the related generalized dimensions. We observe that the dynamics of MCPCA is in all cases characterized with multifractal dynamical response.

Finally, we attempt to analyze the spatial dependence of dynamical correlations $C_{\vec{r}}(t)$. While temporal dependencies seem to follow very similar multiftactal behavior for any $\vec{r}$, their amplitude quickly decays with increasing displacement $\vec{r}$. In order to illustrate this we plot in Fig.~\ref{fig:TACF} time-averaged spatially resolved correlator $\bar{C}(\vec{r}) = \frac{1}{T}\sum_{t=0}^{T-1}C_{\vec{r}}(t)$
and find a quick, possibly isotropic asymptotically exponential decay 
\begin{equation}
    \bar{C}(\vec{r}) \asymp e^{-|\vec{r}|/\xi}.
\end{equation}
Therefore, MCPCA dynamics exhibits a combination of slow dynamics and long-range multifractal order in time with exponential damping in space. This makes it reminiscent of a sort of multiscale glass. 

\section{Summary and conclusion}
We investigated a family of momentum conserving parity check automata (MCPCA) as many-body deterministic dynamical systems defined on extensive bipartite graphs. 
Clear numerical evidence has been provided for a multifractal dynamical response (dynamical structure factor) for models defined on various regular lattices in 2D and 3D. 
 We showed that the response is all-periodic, i.e. we observe subharmonic response at all (dense singular distribution of) rational frequencies, even for a typical fixed trajectory in (thermodynamically) large system.
 In the future, one should attempt to better understand the role of conserved charges and algebraic structures behind fractal dynamics and hierarchical fragmentation in these systems. We conclude that parity check automata
 exhibit previously unseen non-ergodic equilibrium states and
 offer a new mechanism of ergodicity breaking in non-disordered systems 
 fundamentally distinct from integrability. 
 
 Most interestingly,
 the PCA offers direct generalization to
 quantum circuits or quantum cellular automata, where deterministic processes are replaced by parity-check-allowed unitary superpositions. Preliminary investigations indicate possible quantum transitions to/from non-ergodic and multifractal dynamics powered by quantum fluctuations/interference~\cite{PavelOrlov}. On the other hand, replacing unitary processes by stochastic transitions, i.e. investigating stochastic PCA, renders the response $S(\omega)$ smooth and nonsingular. However, as demonstrated in App.~\ref{apx:stoch}, the effect of stochastic noise is continuous: A noise of strength $\gamma$ broadens the delta singularities in $S_{\rm sing}(\omega)$ to width
 $\propto\gamma$ and independent of the system size. This hints that the traces of the phenomenon should be observable in real, noisy systems.

\emph{Acknowledgements}.--- We thank Bruno Bertini, Alexei Kitaev, Katja Klobas, and Roderich Moessner for the fruitful discussions, as well as Pavel Orlov and Cheryne Jonay for their collaboration on a related follow-up project. All power spectra were calculated using the FFTW library~\cite{FFTW05}. The analysis of multifractal dimensions, $\tau$ exponents and singularity spectra followed the methods lined in~\cite{Chhabra89,Martin10}. This research has received funding from the European Union’s Horizon 2020 research and innovation programme under the Marie Sklodowska-Curie grant agreement number 955479, European Research Council (ERC) through Advanced grant QUEST (Grant Agreement No. 101096208), as well as the Slovenian Research and Innovation agency (ARIS) through the Program P1-0402.

\bibliography{bibli.bib}

\newpage
\onecolumngrid
\appendix








\section{Snapshots of the dynamics}
In this appendix, we graphically illustrate MCPCA dynamics on the square and hexagonal lattice enhancing dynamically active regions of particle configurations in real space and demonstrating their very slow deterministic percolation.

\begin{figure}
    \centering
    \includegraphics[width=\linewidth]{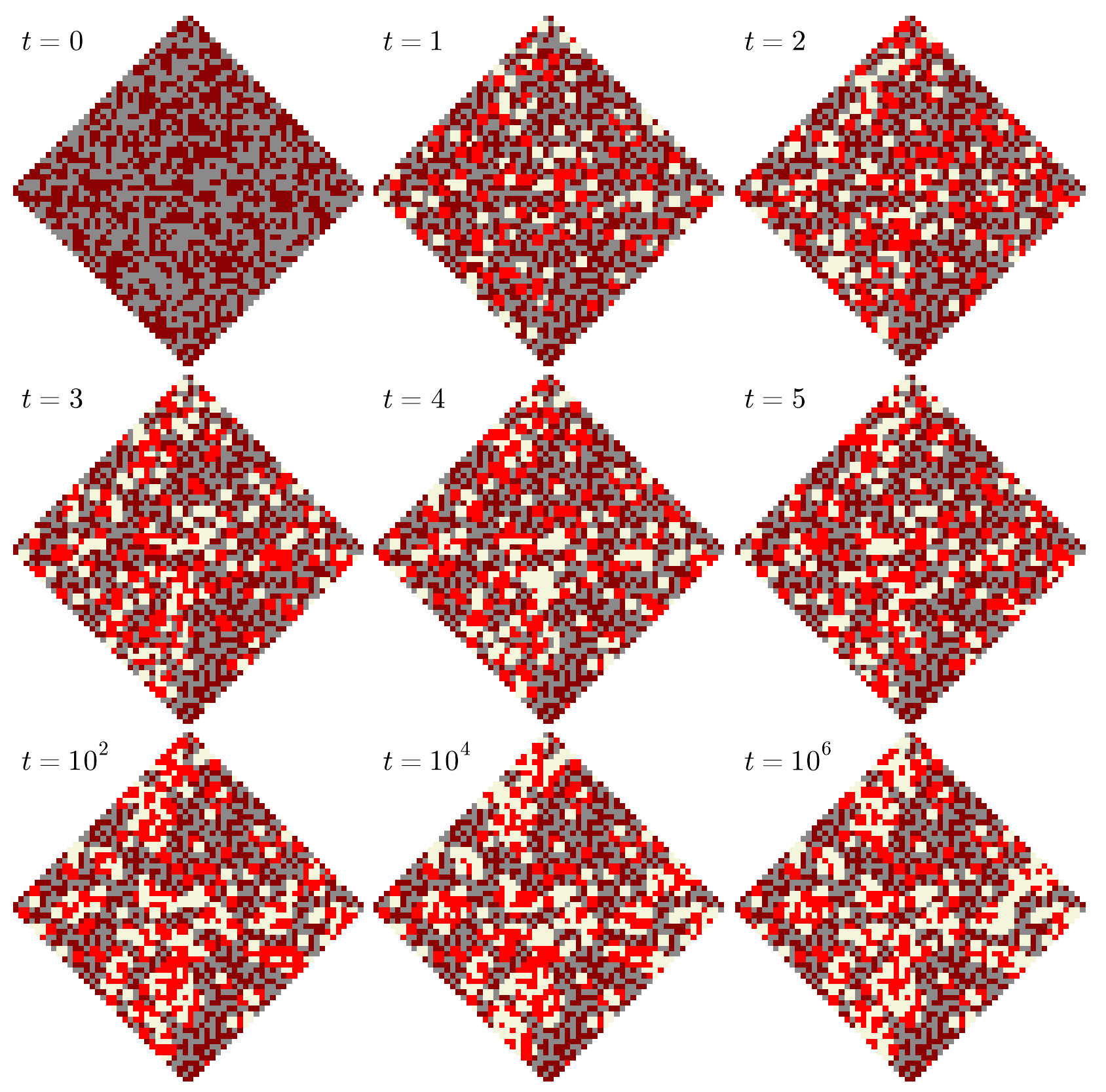}
    \caption{Snapshots of the time evolution (time steps indicated in the figure) of a typical configuration for $32 \times32$ square lattice with periodic boundary conditions and $2\times 32^2$ link degrees of freedom, shown as gray/beige (0) or dark/light red (1) cells. The lighter colors indicate cells which experienced dynamics (at least one flip) up to a given time $t$, while darker colors represent frozen parts of the lattice space.}
    \label{fig:SnapShot}
\end{figure}

To give an insight into space-time dynamics, we plot snapshots of a typical trajectory on a $32\times 32$ square lattice with periodic boundary condition of size $N=2\times 32^2$ (number of edges -- link degrees of freedom) with random initial condition in the random loop-parity sector ($\pi = \pm$). 
We show 9 snapshots of the dynamics in Fig~\ref{fig:SnapShot} up to $t=10^6$ time-steps. We note that the snapshots are rotated by $45^\circ$ with respect to the standard Cartesian grid. The value at each link of the lattice is depicted as a gray (0) or red (1) square. The brighter color indicate cells which have ever flipped from time $0$ to a given time $t$. In this way we get an impression of the slow growth of dynamically active regions of space, which is reminiscent of a slow, deterministic percolation process.

\begin{figure}
    \centering
    \includegraphics[width=\linewidth]{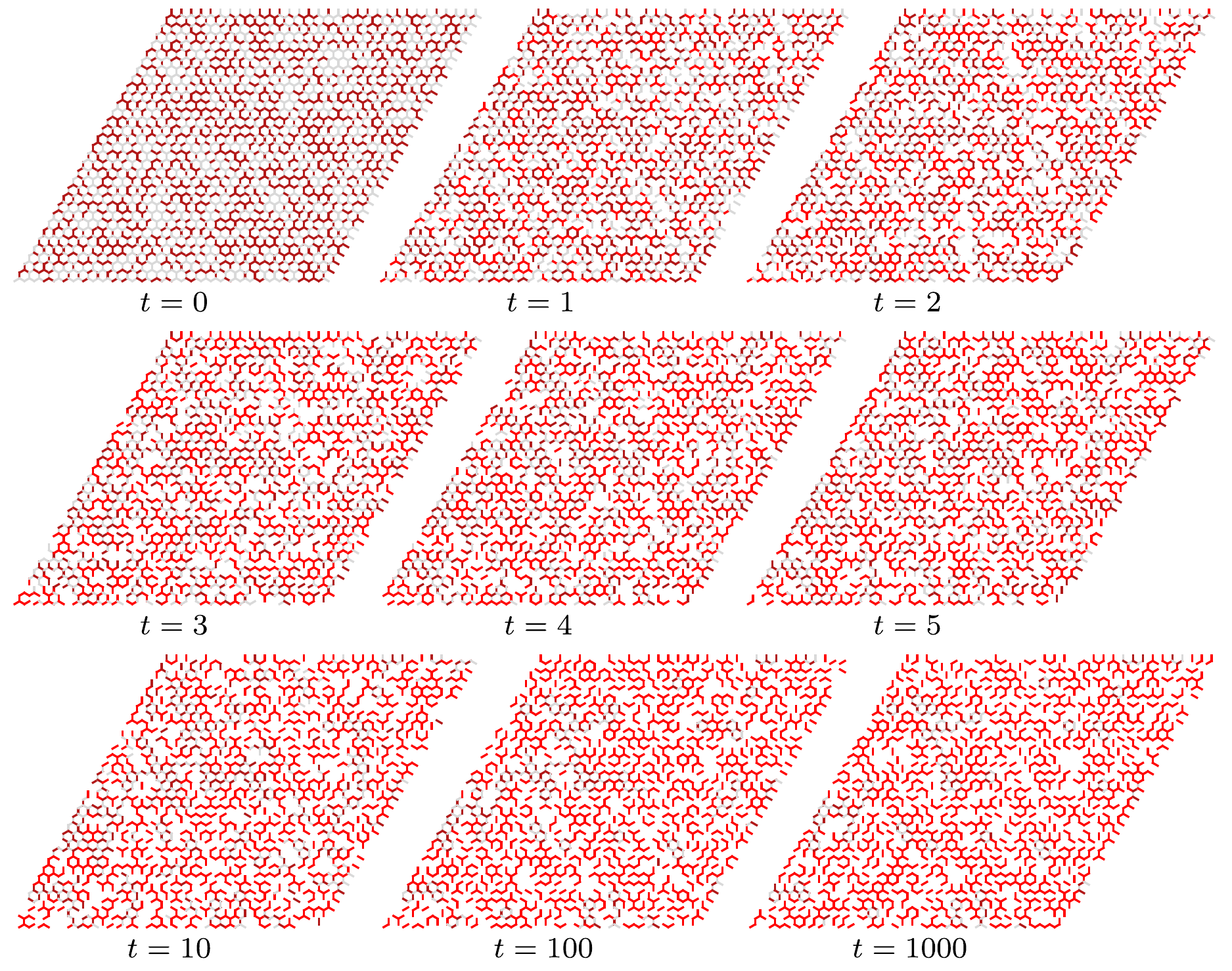}
    \caption{Snapshots of the time evolution (time steps indicated in the figure) of a typical configuration for $32 \times 32$ honeycomb lattice with periodic boundary conditions and similar color coding as in the previous figure (gray/dark red representing inactive cells, white/light red cells representing dynamically active cells). As the cells are associated to links/edges of the honeycomb lattice, we plot them by thick lines.}
    \label{fig:SnapShotHo}
\end{figure}

One might wonder if such slow dynamics as observed in the square lattice rests behind the fundamental reason for multifractality. Hence, we plot in Fig.~\ref{fig:SnapShotHo} similar snapshots for the honeycomb lattice. We see that here, probably due to lower coordination number, the dynamics appears much faster and that the domain of active cells quickly forms a percolating cluster and eventually covers a vast majority of the lattice unlike in the square lattice case. It seems therefore, although the dynamical lattice covering seems an interesting problem of its own, it is not qualitatively connected to multifractality and all-periodic response.

\section{Power spectrum and dynamical response for a global/extensive observable}

\begin{figure}
    \centering
    \includegraphics[width=0.75\linewidth]{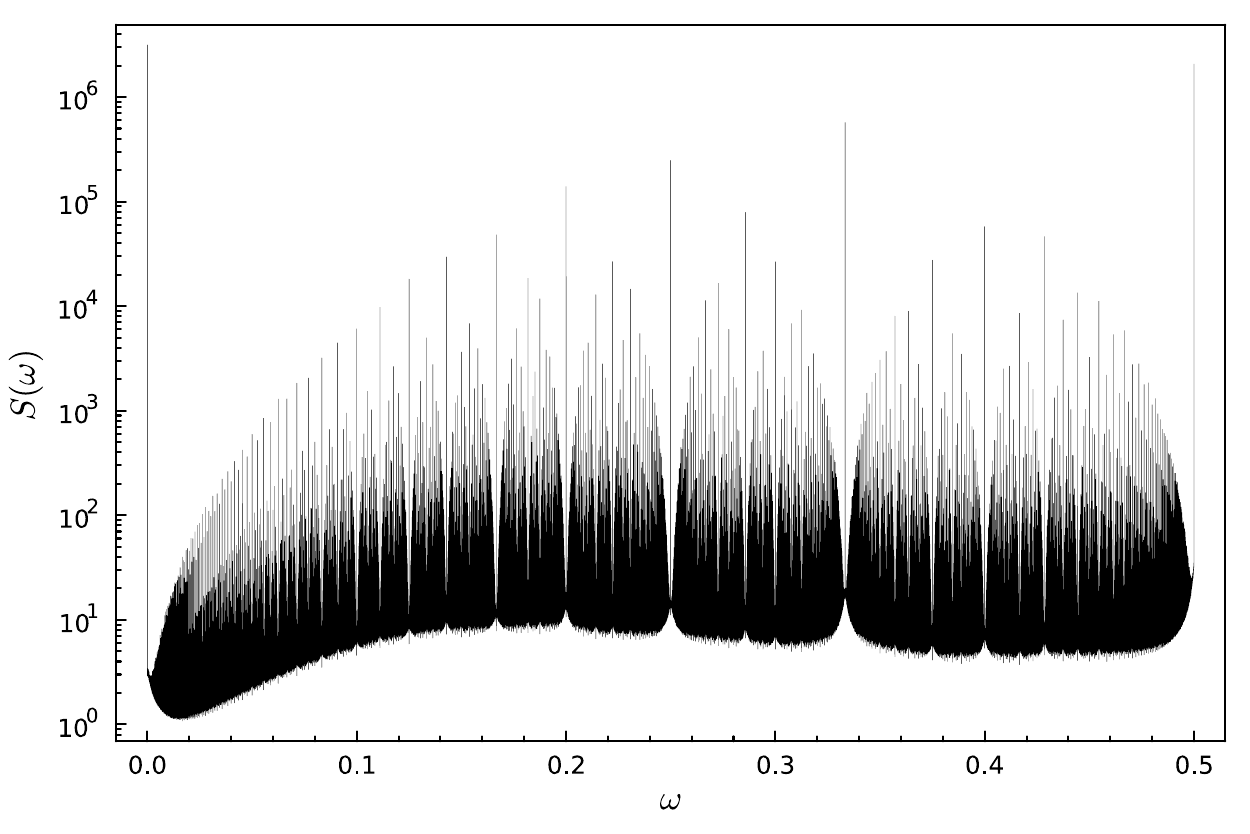}
    \caption{Power spectra of the global (integrated, extensive) density-density correlation functions as a function of the frequency for $16 \times 16$ square lattice ($N=2\cdot 16^2$). The data is calculated for $T=2^{20}$ and for $\mathcal{N} = 3.4\cdot10^5$ all in the random parity sector $\pi=\pm$.}
    \label{fig:spectrum_glob}
\end{figure}

\begin{figure}
    \centering
    \includegraphics[width=0.75\linewidth]{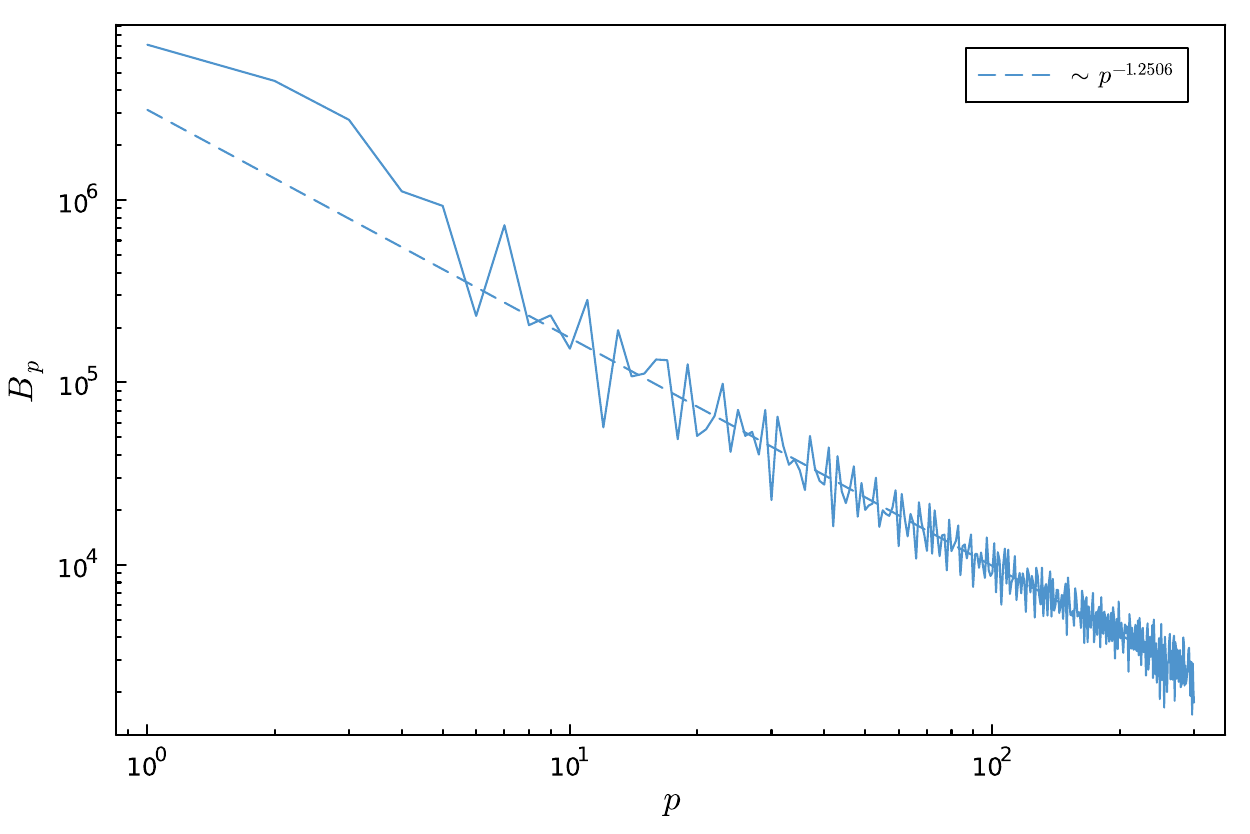}
    \caption{$B_p$ as a function of the period $p$ in log scale for the global density-density correlation function for the square lattice.}
    \label{fig:glob_BP}
\end{figure}

Here we show the dynamical response function -- power spectrum $S(\omega)$ for an example of a global/extensive observable, and demonstrate its similarity to the power spectrum for a local observable. Instead of looking at the density-density correlation function $C_{\vec{r}}(t)$ at $\vec{r}=0$, we now take the autocorrelation function of global/total excitation number 
\begin{equation}
    C(t) = \langle R(\underline{s}) R( \Phi^{(t)}(\underline{s}))  \rangle.
    \end{equation} 
    The extensive, total excitation number (centered around zero) $R(\underline{s})$ is defined as:
\begin{equation}
    R(\underline{s}) =\sum_v \rho_v(\underline{s}).
\end{equation}
Straightforward calculation expresses $C(t)$ as a spatial sum of $C_{\vec{r}}(t)$
\begin{equation}
C(t) =  \sum_{v,v'}
C_{v-v'}(t).
\end{equation}
Fig~\ref{fig:spectrum_glob} shows the power spectrum $S(\omega)$ of the global correlation function $C(t)$ for the square lattice. We observe that the all-periodic multifractal response is (at least qualitatively) indistinguishable from the response for local observable. Furthermore, the power-law decay of $B_p$ for the global observable shown in fig~\ref{fig:glob_BP} is within numerical accuracy identical as for the local observable. These results assure use that the multifractal response is not confined to the local observables, but the behavior is to be expected for
generic local or global observables.

\section{Time average over single long trajectory vs. phase space average}
\label{apx:Time_average}
In the following, we demonstrate an effective ergodicity of dynamical structure factor
 $S(\omega)$ by comparing phase space (initial state) averages with an average over sections of a long, fixed, typical trajectory.

The MCPCA dynamics is clearly non-ergodic
as exemplified by nonzero DC Drude weight $A_{0,1}$ for local and extensive observables. However, the computation of time-nonlocal observables may become `self-averaging' with increasing the system size and time of simulation. For instance, $S(\omega)$ calculated for a fixed,
typical trajectory in a thermodynamically large system can uniformly sample configurations in smaller subsystems, and hence faithfully represent an ensemble (phase space) average. This is indeed suggested by Fig.~\ref{fig:TACF} of the main text, indicating that the spatial correlations decay exponentially.

\begin{figure}
    \centering
    \includegraphics[width=0.85\linewidth]{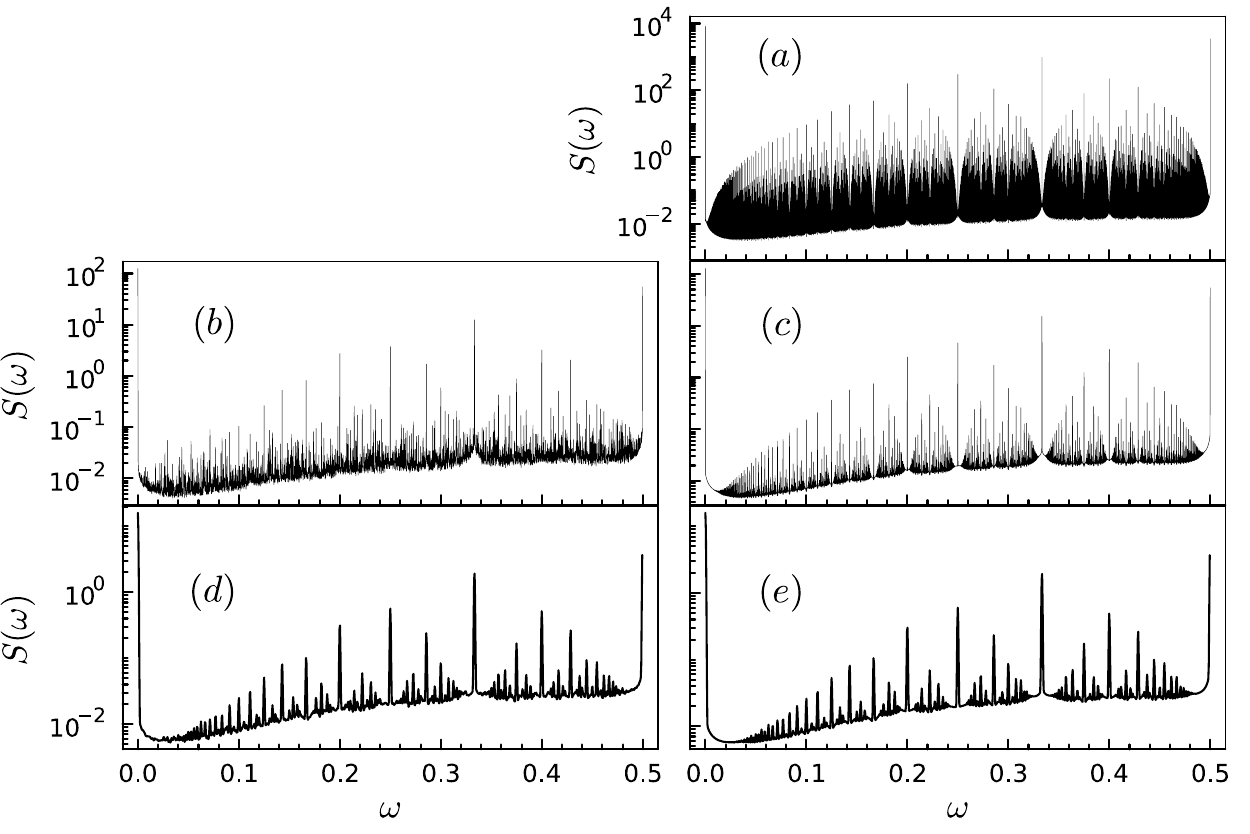}
    \caption{Power spectra $S(\omega)$ for the $L\times L$ periodic square lattice MCPCA, computed either as time averages (left panels, b,d) over a fixed trajectory (averaged over $T_1$ sections, each of length $T_2$, so that the total duration $T=T_1 T_2$) with a random initial condition, or as a phase space average over many ($\mathcal N$) trajectories with independent random initial conditions (right panels, a,c,e).
Panel (a) gives the case $L=16$, $T=2^{20}$, $\mathcal{N} = 0.68\cdot10^{6}$ as also shown in the main text, for comparison. Panels (b) and (c) are for lattice size $L=128$: (b) is for single-trajectory time average with $T_1 =2^{12}$, $T_2=2^{14}$, while (c) is for the phase space average of $\mathcal{N}=2^{12}$ initial conditions and with trajectory duration $T=2^{14}$.
Panels (d) and (e) are for lattices of size $L=512$: (d) is for single-trajectory time average with $T_1=2^{10}$, $T_2=2^{11}$, while (e) is the phase space average of $\mathcal{N}=2^{10}$ initial conditions and with trajectory duration $T=2^{11}$.}
    \label{fig:OneTraj_vs_Statstical}
\end{figure}

 Here we focus on the square lattice as well. We take a lattice of size $L\times L$ with periodic boundary conditions and  and start from a typical, random trajectory with all positive loop-parities. Then we compute $S(\omega)$ as
  a time average over $T_1$ sections of length $T_2$ over a very long trajectory of length $T=T_1 T_2$. We find a clear indication of multifractal power spectrum
  as $L$ and $T_1,T_2$ become sufficiently large, essentially very similar to the phase space average of $S(\omega)$.
  See Fig.~\ref{fig:OneTraj_vs_Statstical} for comparison of results for $L=128 \times 128$ and $L=512 \times 512$ and sufficiently large $T_1,T_2$.

It is clear from figure~\ref{fig:OneTraj_vs_Statstical} that at large enough system sizes, the time averages agree with the average over the phase space. We note that for larger system sizes we have to sacrifice the sharpness of the data by taking shorter final times, this is due to the increase of simulation times as we increase system size. But even at our current resolution, we can see a good agreement for $L=512 \times 512$ in comparison with $L=128 \times 128$ where the amplitudes of peaks are different between the two averages.

\section{Power spectrum of stochastically deformed MCPCA}
\label{apx:stoch}
\begin{figure}
    \centering
    \includegraphics[width=0.75\linewidth]{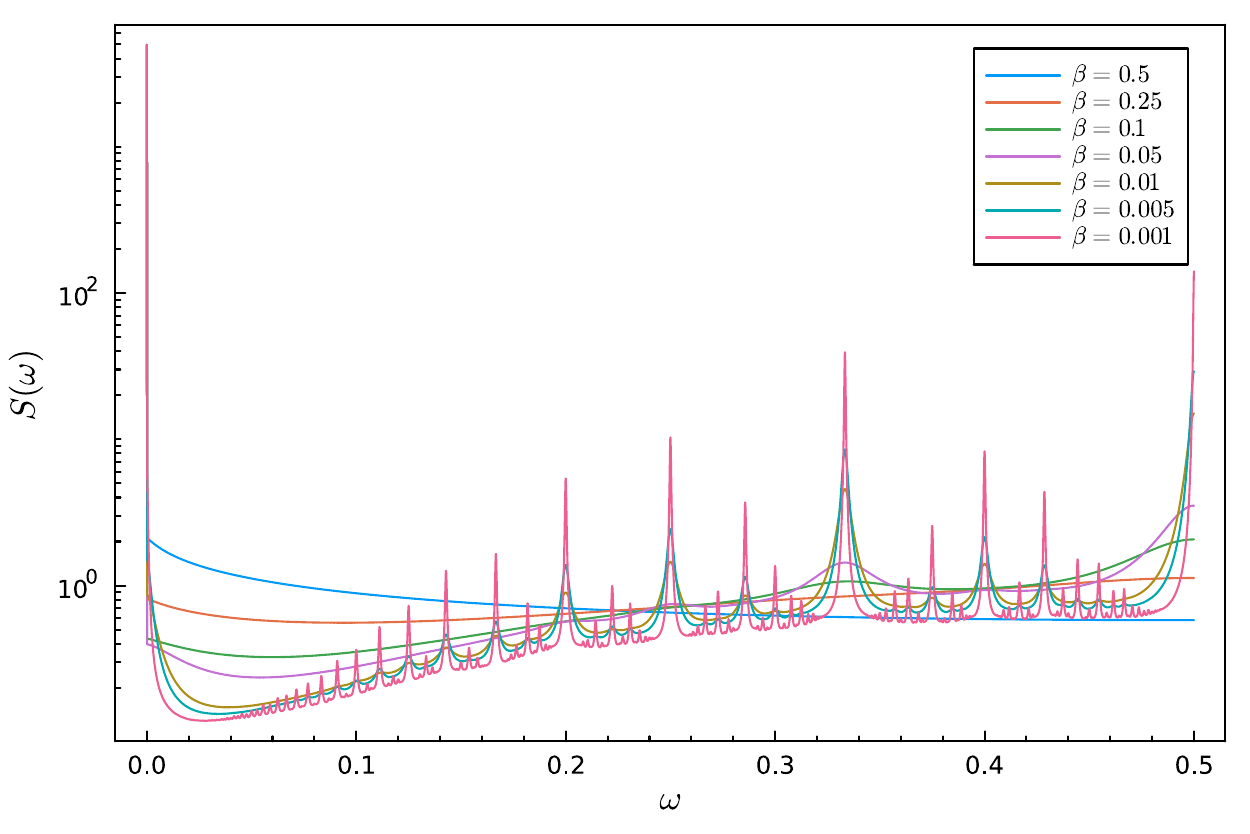}
    \caption{Power spectra $S(\omega)$ of the density-density autocorrelation function $S(\omega)$ for the stochastically deformed PCA on the 
    square lattice of $N = 2\times 16^2$ links. $\beta$ is the probability of stochastically flipping a quadruple of empty/full
    links (\ref{eq:stocchastic_spec}). All the plots are calculated for $T=2^{15}$ with $\mathcal{N} = 2\cdot10^5$ initial conditions in the random parity sector $\pi=\pm$.}
    \label{fig:spectra_sto}
\end{figure}

In this appendix, we define stochastic deformation of MCPCA and demonstrate that multifractality of $S(\omega)$
washes away continuously by increasing the
stochastic jump probabilities.
Specifically, we check the robustness of our main finding, i.e. multifractality of the power spectrum, against a stochastic deformation of MCPCA breaking deterministic and reversible character of the model.
We again consider the square lattice version of the PCA, and stochastically deform two following two rules:
\begin{equation}
	\begin{tikzpicture}[baseline={(current bounding box.center)},every node/.style={inner sep=0,outer sep=0},line cap=rect,scale=0.5]
    \node (n0) at (0,0)[circle,draw,fill,inner sep=1.25pt] {};
    \node (n1) at (1,1) {};
    \node (n2) at (-1,-1) {};
    \node (n3) at (1,-1) {};
    \node (n4) at (-1,1) {};
    \draw[-] (n0) -- (n1);
    \draw[-] (n0) -- (n2);
    \draw[-] (n0) -- (n3);
    \draw[-] (n0) -- (n4);
    \node (A) at (0.5,0.5)[circle,draw,fill=white,inner sep=2.5pt]{};
    \node (B) at (0.5,-0.5)[circle,,draw,fill=white,inner sep=2.5pt]{};
    \node (C) at (-0.5,-0.5)[circle,draw,fill=white,inner sep=2.5pt]{};
    \node (D) at (-0.5,0.5)[circle,,draw,fill=white,inner sep=2.5pt]{};
    \end{tikzpicture}
	\rightarrow \alpha
	\begin{tikzpicture}[baseline={(current bounding box.center)},every node/.style={inner sep=0,outer sep=0},line cap=rect,scale=0.5]
    \node (n0) at (0,0)[circle,draw,fill,inner sep=1.25pt] {};
    \node (n1) at (1,1) {};
    \node (n2) at (-1,-1) {};
    \node (n3) at (1,-1) {};
    \node (n4) at (-1,1) {};
    \draw[-] (n0) -- (n1);
    \draw[-] (n0) -- (n2);
    \draw[-] (n0) -- (n3);
    \draw[-] (n0) -- (n4);
    \node (A) at (0.5,0.5)[circle,draw,fill=white,inner sep=2.5pt]{};
    \node (B) at (0.5,-0.5)[circle,,draw,fill=white,inner sep=2.5pt]{};
    \node (C) at (-0.5,-0.5)[circle,draw,fill=white,inner sep=2.5pt]{};
    \node (D) at (-0.5,0.5)[circle,,draw,fill=white,inner sep=2.5pt]{};
    \end{tikzpicture}
    + \beta
    \begin{tikzpicture}[baseline={(current bounding box.center)},every node/.style={inner sep=0,outer sep=0},line cap=rect,scale=0.5]
    \node (n0) at (0,0)[circle,draw,fill,inner sep=1.25pt] {};
    \node (n1) at (1,1) {};
    \node (n2) at (-1,-1) {};
    \node (n3) at (1,-1) {};
    \node (n4) at (-1,1) {};
    \draw[-] (n0) -- (n1);
    \draw[-] (n0) -- (n2);
    \draw[-] (n0) -- (n3);
    \draw[-] (n0) -- (n4);
    \node (A) at (0.5,0.5)[circle,draw,fill=red,inner sep=2.5pt]{};
    \node (B) at (0.5,-0.5)[circle,,draw,fill=red,inner sep=2.5pt]{};
    \node (C) at (-0.5,-0.5)[circle,draw,fill=red,inner sep=2.5pt]{};
    \node (D) at (-0.5,0.5)[circle,,draw,fill=red,inner sep=2.5pt]{};
    \end{tikzpicture},
    \qquad
    \begin{tikzpicture}[baseline={(current bounding box.center)},every node/.style={inner sep=0,outer sep=0},line cap=rect,scale=0.5]
    \node (n0) at (0,0)[circle,draw,fill,inner sep=1.25pt] {};
    \node (n1) at (1,1) {};
    \node (n2) at (-1,-1) {};
    \node (n3) at (1,-1) {};
    \node (n4) at (-1,1) {};
    \draw[-] (n0) -- (n1);
    \draw[-] (n0) -- (n2);
    \draw[-] (n0) -- (n3);
    \draw[-] (n0) -- (n4);
    \node (A) at (0.5,0.5)[circle,draw,fill=red,inner sep=2.5pt]{};
    \node (B) at (0.5,-0.5)[circle,,draw,fill=red,inner sep=2.5pt]{};
    \node (C) at (-0.5,-0.5)[circle,draw,fill=red,inner sep=2.5pt]{};
    \node (D) at (-0.5,0.5)[circle,,draw,fill=red,inner sep=2.5pt]{};
    \end{tikzpicture}
    \rightarrow \alpha 
    \begin{tikzpicture}[baseline={(current bounding box.center)},every node/.style={inner sep=0,outer sep=0},line cap=rect,scale=0.5]
    \node (n0) at (0,0)[circle,draw,fill,inner sep=1.25pt] {};
    \node (n1) at (1,1) {};
    \node (n2) at (-1,-1) {};
    \node (n3) at (1,-1) {};
    \node (n4) at (-1,1) {};
    \draw[-] (n0) -- (n1);
    \draw[-] (n0) -- (n2);
    \draw[-] (n0) -- (n3);
    \draw[-] (n0) -- (n4);
    \node (A) at (0.5,0.5)[circle,draw,fill=red,inner sep=2.5pt]{};
    \node (B) at (0.5,-0.5)[circle,,draw,fill=red,inner sep=2.5pt]{};
    \node (C) at (-0.5,-0.5)[circle,draw,fill=red,inner sep=2.5pt]{};
    \node (D) at (-0.5,0.5)[circle,,draw,fill=red,inner sep=2.5pt]{};
    \end{tikzpicture}
    + \beta
    \begin{tikzpicture}[baseline={(current bounding box.center)},every node/.style={inner sep=0,outer sep=0},line cap=rect,scale=0.5]
    \node (n0) at (0,0)[circle,draw,fill,inner sep=1.25pt] {};
    \node (n1) at (1,1) {};
    \node (n2) at (-1,-1) {};
    \node (n3) at (1,-1) {};
    \node (n4) at (-1,1) {};
    \draw[-] (n0) -- (n1);
    \draw[-] (n0) -- (n2);
    \draw[-] (n0) -- (n3);
    \draw[-] (n0) -- (n4);
    \node (A) at (0.5,0.5)[circle,draw,fill=white,inner sep=2.5pt]{};
    \node (B) at (0.5,-0.5)[circle,,draw,fill=white,inner sep=2.5pt]{};
    \node (C) at (-0.5,-0.5)[circle,draw,fill=white,inner sep=2.5pt]{};
    \node (D) at (-0.5,0.5)[circle,,draw,fill=white,inner sep=2.5pt]{};
    \end{tikzpicture}
    \label{eq:stocchastic_spec}
\end{equation}
where $\beta = 1-\alpha \in (0,1)$ is the
probability that all empty configuration
$(0000)$ around the vertex flips to all full $(1111)$, or vice versa.
Note that $\beta=0$ corresponds to a deterministic MCPCA investigated previously,
while $\beta=1$ corresponds to trivial dynamics (flipping all cells at each half-time-step). Such stochastically deformed MCPCA becomes now a bistochastic Markov chain with the same local constants of motion (loop-parities, and Neel loops).
Nevertheless, as we demonstrate in 
Fig~\ref{fig:spectra_sto} the $\delta-$singularities (AC Drude peaks) in $S(\omega)$ are broadened proportionally to $\beta$ so the power spectrum becomes smooth for any finite value of $\beta\in(0,1)$.

\section{PCA with broken momentum conservation}

Following up on the question of robustness from the last appendix, we will now investigate the effect of the breaking of momenta conservation while keeping the parity check and the reversibility intact.
In fact, we consider discrete deformations of our PCA update rule $\Phi_v$ which map $(s_1,s_2\ldots s_n)\to(\bar{s}_1,\bar{s}_2\ldots\bar{s}_n)$
for all but one pair of mutually negated configurations which both get mapped to itself
$(s'_1,s'_2\ldots s'_n)\to (s'_1,s'_2\ldots s'_n)$,
$(\bar{s}'_1,\bar{s}'_2\ldots \bar{s}'_n)\to (\bar{s}'_1,\bar{s}'_2\ldots \bar{s}'_n)$.
We break momentum conservation if we choose $(s'_1,s'_2\ldots s'_n)$ different from $(00\ldots 0)$ or $(11\ldots 1)$.

We consider the following explicit deformation with a complete set of rules:
\begin{eqnarray}
	\begin{tikzpicture}[baseline={(current bounding box.center)},every node/.style={inner sep=0,outer sep=0},line cap=rect,scale=0.5]
    \node (n0) at (0,0)[circle,draw,fill,inner sep=1.25pt] {};
    \node (n1) at (1,1) {};
    \node (n2) at (-1,-1) {};
    \node (n3) at (1,-1) {};
    \node (n4) at (-1,1) {};
    \draw[-] (n0) -- (n1);
    \draw[-] (n0) -- (n2);
    \draw[-] (n0) -- (n3);
    \draw[-] (n0) -- (n4);
    \node (A) at (0.5,0.5)[circle,draw,fill=white,inner sep=2.5pt]{};
    \node (B) at (0.5,-0.5)[circle,,draw,fill=white,inner sep=2.5pt]{};
    \node (C) at (-0.5,-0.5)[circle,draw,fill=white,inner sep=2.5pt]{};
    \node (D) at (-0.5,0.5)[circle,,draw,fill=white,inner sep=2.5pt]{};
    \end{tikzpicture}
	\leftrightarrow
	\begin{tikzpicture}[baseline={(current bounding box.center)},every node/.style={inner sep=0,outer sep=0},line cap=rect,scale=0.5]
    \node (n0) at (0,0)[circle,draw,fill,inner sep=1.25pt] {};
    \node (n1) at (1,1) {};
    \node (n2) at (-1,-1) {};
    \node (n3) at (1,-1) {};
    \node (n4) at (-1,1) {};
    \draw[-] (n0) -- (n1);
    \draw[-] (n0) -- (n2);
    \draw[-] (n0) -- (n3);
    \draw[-] (n0) -- (n4);
    \node (A) at (0.5,0.5)[circle,draw,fill=red,inner sep=2.5pt]{};
    \node (B) at (0.5,-0.5)[circle,,draw,fill=red,inner sep=2.5pt]{};
    \node (C) at (-0.5,-0.5)[circle,draw,fill=red,inner sep=2.5pt]{};
    \node (D) at (-0.5,0.5)[circle,,draw,fill=red,inner sep=2.5pt]{};
    \end{tikzpicture},
    &\quad&
	\begin{tikzpicture}[baseline={(current bounding box.center)},every node/.style={inner sep=0,outer sep=0},line cap=rect,scale=0.5]
    \node (n0) at (0,0)[circle,draw,fill,inner sep=1.25pt] {};
    \node (n1) at (1,1) {};
    \node (n2) at (-1,-1) {};
    \node (n3) at (1,-1) {};
    \node (n4) at (-1,1) {};
    \draw[-] (n0) -- (n1);
    \draw[-] (n0) -- (n2);
    \draw[-] (n0) -- (n3);
    \draw[-] (n0) -- (n4);
    \node (A) at (0.5,0.5)[circle,draw,fill=red,inner sep=2.5pt]{};
    \node (B) at (0.5,-0.5)[circle,,draw,fill=white,inner sep=2.5pt]{};
    \node (C) at (-0.5,-0.5)[circle,draw,fill=white,inner sep=2.5pt]{};
    \node (D) at (-0.5,0.5)[circle,,draw,fill=white,inner sep=2.5pt]{};
    \end{tikzpicture}
	\leftrightarrow
	\begin{tikzpicture}[baseline={(current bounding box.center)},every node/.style={inner sep=0,outer sep=0},line cap=rect,scale=0.5]
    \node (n0) at (0,0)[circle,draw,fill,inner sep=1.25pt] {};
    \node (n1) at (1,1) {};
    \node (n2) at (-1,-1) {};
    \node (n3) at (1,-1) {};
    \node (n4) at (-1,1) {};
    \draw[-] (n0) -- (n1);
    \draw[-] (n0) -- (n2);
    \draw[-] (n0) -- (n3);
    \draw[-] (n0) -- (n4);
    \node (A) at (0.5,0.5)[circle,draw,fill=red,inner sep=2.5pt]{};
    \node (B) at (0.5,-0.5)[circle,,draw,fill=white,inner sep=2.5pt]{};
    \node (C) at (-0.5,-0.5)[circle,draw,fill=white,inner sep=2.5pt]{};
    \node (D) at (-0.5,0.5)[circle,,draw,fill=white,inner sep=2.5pt]{};
    \end{tikzpicture},
    \quad
    \begin{tikzpicture}[baseline={(current bounding box.center)},every node/.style={inner sep=0,outer sep=0},line cap=rect,scale=0.5]
    \node (n0) at (0,0)[circle,draw,fill,inner sep=1.25pt] {};
    \node (n1) at (1,1) {};
    \node (n2) at (-1,-1) {};
    \node (n3) at (1,-1) {};
    \node (n4) at (-1,1) {};
    \draw[-] (n0) -- (n1);
    \draw[-] (n0) -- (n2);
    \draw[-] (n0) -- (n3);
    \draw[-] (n0) -- (n4);
    \node (A) at (0.5,0.5)[circle,draw,fill=white,inner sep=2.5pt]{};
    \node (B) at (0.5,-0.5)[circle,,draw,fill=red,inner sep=2.5pt]{};
    \node (C) at (-0.5,-0.5)[circle,draw,fill=red,inner sep=2.5pt]{};
    \node (D) at (-0.5,0.5)[circle,,draw,fill=red,inner sep=2.5pt]{};
    \end{tikzpicture}
	\leftrightarrow
	\begin{tikzpicture}[baseline={(current bounding box.center)},every node/.style={inner sep=0,outer sep=0},line cap=rect,scale=0.5]
    \node (n0) at (0,0)[circle,draw,fill,inner sep=1.25pt] {};
    \node (n1) at (1,1) {};
    \node (n2) at (-1,-1) {};
    \node (n3) at (1,-1) {};
    \node (n4) at (-1,1) {};
    \draw[-] (n0) -- (n1);
    \draw[-] (n0) -- (n2);
    \draw[-] (n0) -- (n3);
    \draw[-] (n0) -- (n4);
    \node (A) at (0.5,0.5)[circle,draw,fill=white,inner sep=2.5pt]{};
    \node (B) at (0.5,-0.5)[circle,,draw,fill=red,inner sep=2.5pt]{};
    \node (C) at (-0.5,-0.5)[circle,draw,fill=red,inner sep=2.5pt]{};
    \node (D) at (-0.5,0.5)[circle,,draw,fill=red,inner sep=2.5pt]{};
    \end{tikzpicture}
    \nonumber \\
    \begin{tikzpicture}[baseline={(current bounding box.center)},every node/.style={inner sep=0,outer sep=0},line cap=rect,scale=0.5]
    \node (n0) at (0,0)[circle,draw,fill,inner sep=1.25pt] {};
    \node (n1) at (1,1) {};
    \node (n2) at (-1,-1) {};
    \node (n3) at (1,-1) {};
    \node (n4) at (-1,1) {};
    \draw[-] (n0) -- (n1);
    \draw[-] (n0) -- (n2);
    \draw[-] (n0) -- (n3);
    \draw[-] (n0) -- (n4);
    \node (A) at (0.5,0.5)[circle,draw,fill=white,inner sep=2.5pt]{};
    \node (B) at (0.5,-0.5)[circle,,draw,fill=red,inner sep=2.5pt]{};
    \node (C) at (-0.5,-0.5)[circle,draw,fill=white,inner sep=2.5pt]{};
    \node (D) at (-0.5,0.5)[circle,,draw,fill=white,inner sep=2.5pt]{};
    \end{tikzpicture}
	\leftrightarrow
	\begin{tikzpicture}[baseline={(current bounding box.center)},every node/.style={inner sep=0,outer sep=0},line cap=rect,scale=0.5]
    \node (n0) at (0,0)[circle,draw,fill,inner sep=1.25pt] {};
    \node (n1) at (1,1) {};
    \node (n2) at (-1,-1) {};
    \node (n3) at (1,-1) {};
    \node (n4) at (-1,1) {};
    \draw[-] (n0) -- (n1);
    \draw[-] (n0) -- (n2);
    \draw[-] (n0) -- (n3);
    \draw[-] (n0) -- (n4);
    \node (A) at (0.5,0.5)[circle,draw,fill=red,inner sep=2.5pt]{};
    \node (B) at (0.5,-0.5)[circle,,draw,fill=white,inner sep=2.5pt]{};
    \node (C) at (-0.5,-0.5)[circle,draw,fill=red,inner sep=2.5pt]{};
    \node (D) at (-0.5,0.5)[circle,,draw,fill=red,inner sep=2.5pt]{};
    \end{tikzpicture},
    &\quad&
	\begin{tikzpicture}[baseline={(current bounding box.center)},every node/.style={inner sep=0,outer sep=0},line cap=rect,scale=0.5]
    \node (n0) at (0,0)[circle,draw,fill,inner sep=1.25pt] {};
    \node (n1) at (1,1) {};
    \node (n2) at (-1,-1) {};
    \node (n3) at (1,-1) {};
    \node (n4) at (-1,1) {};
    \draw[-] (n0) -- (n1);
    \draw[-] (n0) -- (n2);
    \draw[-] (n0) -- (n3);
    \draw[-] (n0) -- (n4);
    \node (A) at (0.5,0.5)[circle,draw,fill=white,inner sep=2.5pt]{};
    \node (B) at (0.5,-0.5)[circle,,draw,fill=white,inner sep=2.5pt]{};
    \node (C) at (-0.5,-0.5)[circle,draw,fill=red,inner sep=2.5pt]{};
    \node (D) at (-0.5,0.5)[circle,,draw,fill=white,inner sep=2.5pt]{};
    \end{tikzpicture}
	\leftrightarrow
	\begin{tikzpicture}[baseline={(current bounding box.center)},every node/.style={inner sep=0,outer sep=0},line cap=rect,scale=0.5]
    \node (n0) at (0,0)[circle,draw,fill,inner sep=1.25pt] {};
    \node (n1) at (1,1) {};
    \node (n2) at (-1,-1) {};
    \node (n3) at (1,-1) {};
    \node (n4) at (-1,1) {};
    \draw[-] (n0) -- (n1);
    \draw[-] (n0) -- (n2);
    \draw[-] (n0) -- (n3);
    \draw[-] (n0) -- (n4);
    \node (A) at (0.5,0.5)[circle,draw,fill=red,inner sep=2.5pt]{};
    \node (B) at (0.5,-0.5)[circle,,draw,fill=red,inner sep=2.5pt]{};
    \node (C) at (-0.5,-0.5)[circle,draw,fill=white,inner sep=2.5pt]{};
    \node (D) at (-0.5,0.5)[circle,,draw,fill=red,inner sep=2.5pt]{};
    \end{tikzpicture},
    \quad
	\begin{tikzpicture}[baseline={(current bounding box.center)},every node/.style={inner sep=0,outer sep=0},line cap=rect,scale=0.5]
    \node (n0) at (0,0)[circle,draw,fill,inner sep=1.25pt] {};
    \node (n1) at (1,1) {};
    \node (n2) at (-1,-1) {};
    \node (n3) at (1,-1) {};
    \node (n4) at (-1,1) {};
    \draw[-] (n0) -- (n1);
    \draw[-] (n0) -- (n2);
    \draw[-] (n0) -- (n3);
    \draw[-] (n0) -- (n4);
    \node (A) at (0.5,0.5)[circle,draw,fill=white,inner sep=2.5pt]{};
    \node (B) at (0.5,-0.5)[circle,,draw,fill=white,inner sep=2.5pt]{};
    \node (C) at (-0.5,-0.5)[circle,draw,fill=white,inner sep=2.5pt]{};
    \node (D) at (-0.5,0.5)[circle,,draw,fill=red,inner sep=2.5pt]{};
    \end{tikzpicture}
	\leftrightarrow
	\begin{tikzpicture}[baseline={(current bounding box.center)},every node/.style={inner sep=0,outer sep=0},line cap=rect,scale=0.5]
    \node (n0) at (0,0)[circle,draw,fill,inner sep=1.25pt] {};
    \node (n1) at (1,1) {};
    \node (n2) at (-1,-1) {};
    \node (n3) at (1,-1) {};
    \node (n4) at (-1,1) {};
    \draw[-] (n0) -- (n1);
    \draw[-] (n0) -- (n2);
    \draw[-] (n0) -- (n3);
    \draw[-] (n0) -- (n4);
    \node (A) at (0.5,0.5)[circle,draw,fill=red,inner sep=2.5pt]{};
    \node (B) at (0.5,-0.5)[circle,,draw,fill=red,inner sep=2.5pt]{};
    \node (C) at (-0.5,-0.5)[circle,draw,fill=red,inner sep=2.5pt]{};
    \node (D) at (-0.5,0.5)[circle,,draw,fill=white,inner sep=2.5pt]{};
    \end{tikzpicture}
    \nonumber \\
    \begin{tikzpicture}[baseline={(current bounding box.center)},every node/.style={inner sep=0,outer sep=0},line cap=rect,scale=0.5]
    \node (n0) at (0,0)[circle,draw,fill,inner sep=1.25pt] {};
    \node (n1) at (1,1) {};
    \node (n2) at (-1,-1) {};
    \node (n3) at (1,-1) {};
    \node (n4) at (-1,1) {};
    \draw[-] (n0) -- (n1);
    \draw[-] (n0) -- (n2);
    \draw[-] (n0) -- (n3);
    \draw[-] (n0) -- (n4);
    \node (A) at (0.5,0.5)[circle,draw,fill=white,inner sep=2.5pt]{};
    \node (B) at (0.5,-0.5)[circle,,draw,fill=red,inner sep=2.5pt]{};
    \node (C) at (-0.5,-0.5)[circle,draw,fill=white,inner sep=2.5pt]{};
    \node (D) at (-0.5,0.5)[circle,,draw,fill=red,inner sep=2.5pt]{};
    \end{tikzpicture}
	\leftrightarrow
	\begin{tikzpicture}[baseline={(current bounding box.center)},every node/.style={inner sep=0,outer sep=0},line cap=rect,scale=0.5]
    \node (n0) at (0,0)[circle,draw,fill,inner sep=1.25pt] {};
    \node (n1) at (1,1) {};
    \node (n2) at (-1,-1) {};
    \node (n3) at (1,-1) {};
    \node (n4) at (-1,1) {};
    \draw[-] (n0) -- (n1);
    \draw[-] (n0) -- (n2);
    \draw[-] (n0) -- (n3);
    \draw[-] (n0) -- (n4);
    \node (A) at (0.5,0.5)[circle,draw,fill=red,inner sep=2.5pt]{};
    \node (B) at (0.5,-0.5)[circle,,draw,fill=white,inner sep=2.5pt]{};
    \node (C) at (-0.5,-0.5)[circle,draw,fill=red,inner sep=2.5pt]{};
    \node (D) at (-0.5,0.5)[circle,,draw,fill=white,inner sep=2.5pt]{};
    \end{tikzpicture},
    &\quad&
	\begin{tikzpicture}[baseline={(current bounding box.center)},every node/.style={inner sep=0,outer sep=0},line cap=rect,scale=0.5]
    \node (n0) at (0,0)[circle,draw,fill,inner sep=1.25pt] {};
    \node (n1) at (1,1) {};
    \node (n2) at (-1,-1) {};
    \node (n3) at (1,-1) {};
    \node (n4) at (-1,1) {};
    \draw[-] (n0) -- (n1);
    \draw[-] (n0) -- (n2);
    \draw[-] (n0) -- (n3);
    \draw[-] (n0) -- (n4);
    \node (A) at (0.5,0.5)[circle,draw,fill=white,inner sep=2.5pt]{};
    \node (B) at (0.5,-0.5)[circle,,draw,fill=white,inner sep=2.5pt]{};
    \node (C) at (-0.5,-0.5)[circle,draw,fill=red,inner sep=2.5pt]{};
    \node (D) at (-0.5,0.5)[circle,,draw,fill=red,inner sep=2.5pt]{};
    \end{tikzpicture}
	\leftrightarrow
	\begin{tikzpicture}[baseline={(current bounding box.center)},every node/.style={inner sep=0,outer sep=0},line cap=rect,scale=0.5]
    \node (n0) at (0,0)[circle,draw,fill,inner sep=1.25pt] {};
    \node (n1) at (1,1) {};
    \node (n2) at (-1,-1) {};
    \node (n3) at (1,-1) {};
    \node (n4) at (-1,1) {};
    \draw[-] (n0) -- (n1);
    \draw[-] (n0) -- (n2);
    \draw[-] (n0) -- (n3);
    \draw[-] (n0) -- (n4);
    \node (A) at (0.5,0.5)[circle,draw,fill=red,inner sep=2.5pt]{};
    \node (B) at (0.5,-0.5)[circle,,draw,fill=red,inner sep=2.5pt]{};
    \node (C) at (-0.5,-0.5)[circle,draw,fill=white,inner sep=2.5pt]{};
    \node (D) at (-0.5,0.5)[circle,,draw,fill=white,inner sep=2.5pt]{};
    \end{tikzpicture},
    \quad
	\begin{tikzpicture}[baseline={(current bounding box.center)},every node/.style={inner sep=0,outer sep=0},line cap=rect,scale=0.5]
    \node (n0) at (0,0)[circle,draw,fill,inner sep=1.25pt] {};
    \node (n1) at (1,1) {};
    \node (n2) at (-1,-1) {};
    \node (n3) at (1,-1) {};
    \node (n4) at (-1,1) {};
    \draw[-] (n0) -- (n1);
    \draw[-] (n0) -- (n2);
    \draw[-] (n0) -- (n3);
    \draw[-] (n0) -- (n4);
    \node (A) at (0.5,0.5)[circle,draw,fill=red,inner sep=2.5pt]{};
    \node (B) at (0.5,-0.5)[circle,,draw,fill=white,inner sep=2.5pt]{};
    \node (C) at (-0.5,-0.5)[circle,draw,fill=white,inner sep=2.5pt]{};
    \node (D) at (-0.5,0.5)[circle,,draw,fill=red,inner sep=2.5pt]{};
    \end{tikzpicture}
	\leftrightarrow
	\begin{tikzpicture}[baseline={(current bounding box.center)},every node/.style={inner sep=0,outer sep=0},line cap=rect,scale=0.5]
    \node (n0) at (0,0)[circle,draw,fill,inner sep=1.25pt] {};
    \node (n1) at (1,1) {};
    \node (n2) at (-1,-1) {};
    \node (n3) at (1,-1) {};
    \node (n4) at (-1,1) {};
    \draw[-] (n0) -- (n1);
    \draw[-] (n0) -- (n2);
    \draw[-] (n0) -- (n3);
    \draw[-] (n0) -- (n4);
    \node (A) at (0.5,0.5)[circle,draw,fill=white,inner sep=2.5pt]{};
    \node (B) at (0.5,-0.5)[circle,,draw,fill=red,inner sep=2.5pt]{};
    \node (C) at (-0.5,-0.5)[circle,draw,fill=red,inner sep=2.5pt]{};
    \node (D) at (-0.5,0.5)[circle,,draw,fill=white,inner sep=2.5pt]{};
    \end{tikzpicture}
    \label{S3}
\end{eqnarray}

\begin{figure}
    \centering
    \includegraphics[width=0.75\linewidth]{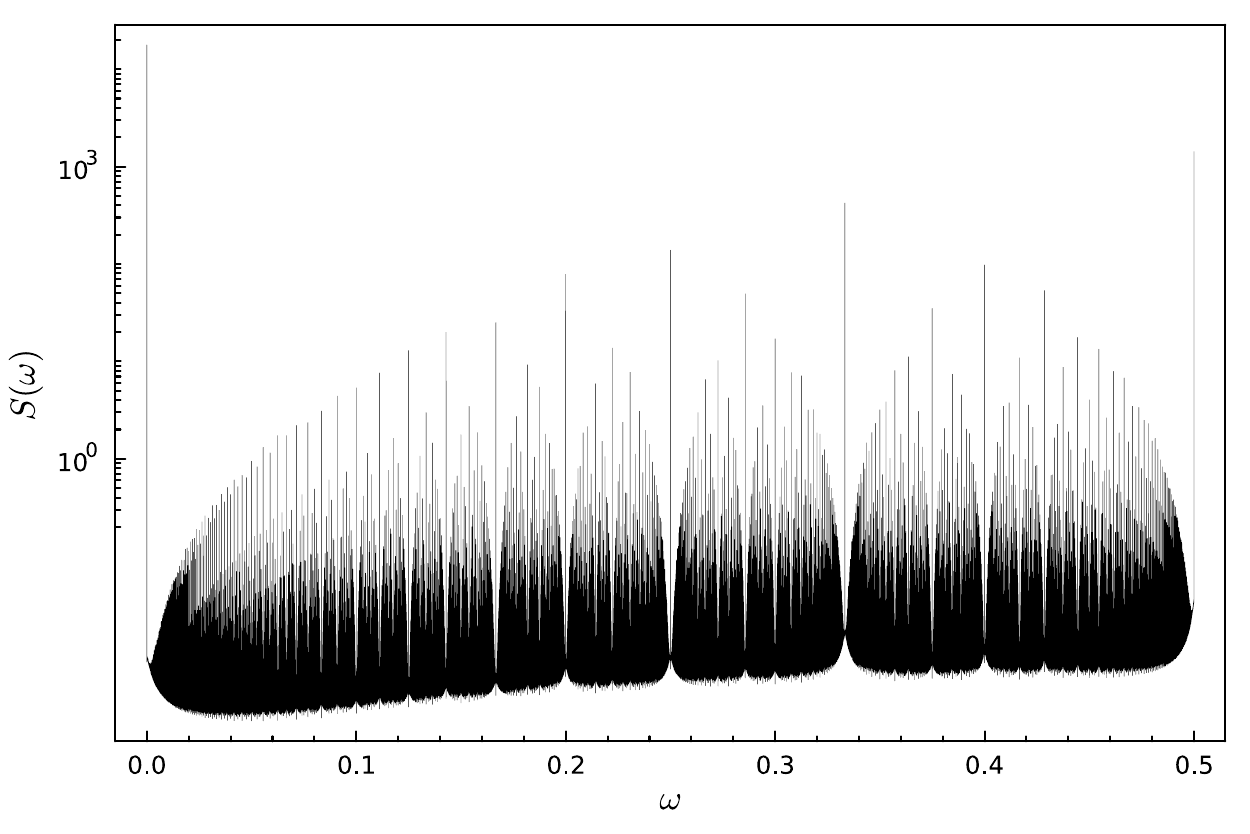}
    \caption{Power spectrum $S(\omega)$ of the density-density autocorrelation function for the square lattice with size $N=2\cdot 16^2$ and for $T=2^{20}$,
    for the PCA with broken momentum conservation (\ref{S3}). The data is averaged over $\mathcal{N}=4.7\cdot10^5$ random initial conditions in the random loop-parity sector $\pi=\pm$.}
    \label{fig:spectrum_nomomenta}
\end{figure}

\begin{figure}
    \centering
    \includegraphics[width=0.75\linewidth]{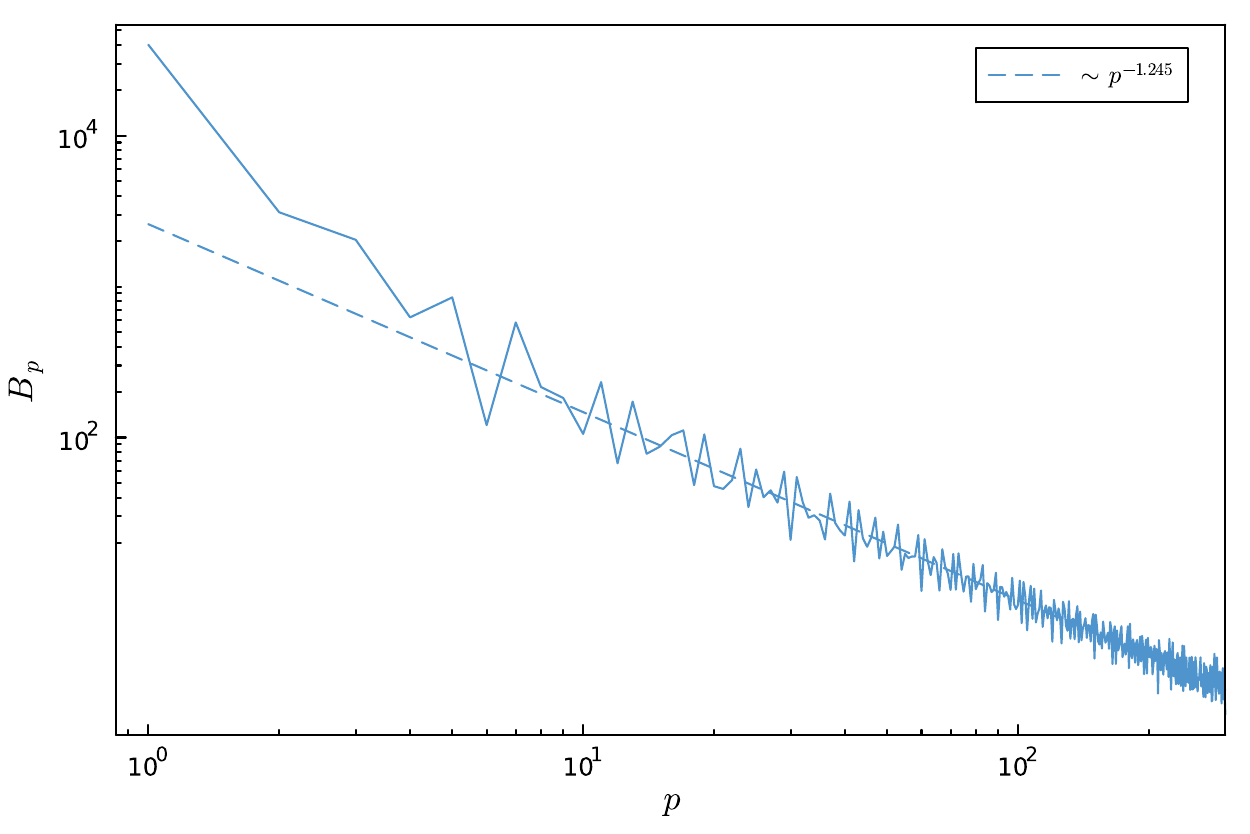}
    \caption{Spectral weight $B_p$ as a function of the period $p$ in log scale for the square lattice with broken momentum conservation.}
    \label{fig:nomomenta_BP}
\end{figure}

In fig~\ref{fig:spectrum_nomomenta} we show the corresponding power spectra,
and demonstrating that it exhibits very similar (if not identical) multifractal spectrum as the MCPCA considered in the main text.
Further investigation of the data in figure~\ref{fig:nomomenta_BP} shows that periodic weights $B_p$ exhibits a power law decay with a similar, if not identical exponent $\mu\approx 1.25$ as that of the original MCPCA on the square lattice in the random parity sector. 
We also verified that other momentum conservation breaking deformations described above yield essentially identical power spectrum.

Note however that we can also build more complex deformations of MCPCA but still satisfying the parity-check constraint, say such that two (or more) distinct pairs of configurations, 
$(s^{(p)}_1,s^{(p)}_2\ldots s^{(p)}_n)$ (and the corresponding negation) map back to themselves, for $p=1,2\ldots P$, $P\ge 2$. Quick inspection of power spectra fur such PCA models show even distinctly richer structures, depending also on the geometry and dimensionality of the lattice (not shown here). Say for the square lattice  and $P=2$,
$S(\omega)$ seem to converge to a continuous but non-smooth function with derivatives experiencing finite jumps at rational frequencies which get smaller with increasing denominators, so a graph of the derivative $S'(\omega)$ may become multifractal.

Finally, what are the effects of deforming the parity-check constraint, i.e. breaking the loop-parity conservation? An old example of a similar cellular automaton with momenta conservation and reversible dynamics is the HPP model~\cite{Hardy73,Hardy76}, which is defined on the square lattice. In addition, this model also conserves the particle number. Clearly, the power spectrum $S(\omega)$ of the HPP model is not multifractal, in fact it has continuous derivatives at all orders (it is a real analytic function).
We can then postulate that the all-periodic multifractal response is coming as a result of the interplay between deterministic reversibility and parity check constraint.

\section{The calculation of generalized dimensions and $\tau(q)$ exponents
of the multifractal power spectrum}
\label{apx:details_tau}

\begin{figure}
    \centering
    \includegraphics[width=0.75\linewidth]{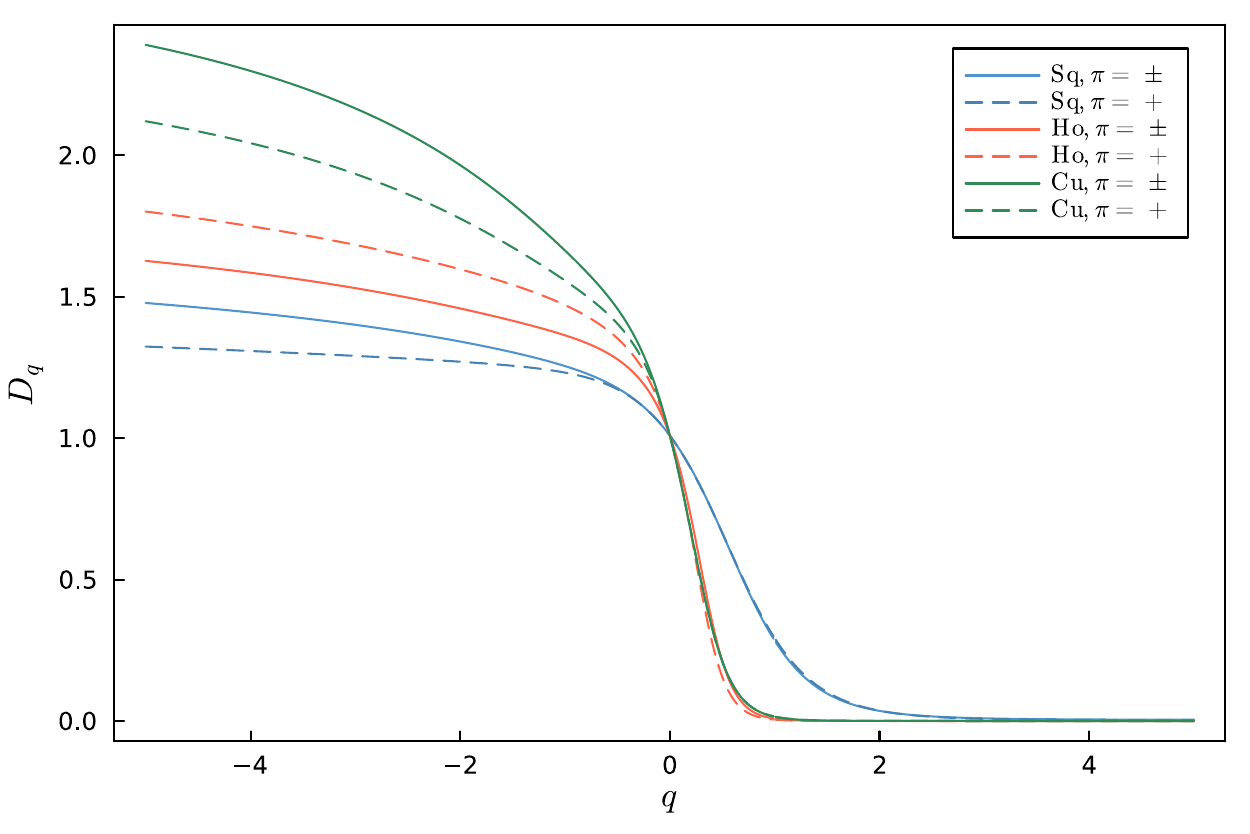}
    \caption{The generalized dimension $D_q$ as a function of $q$ for different lattices, Sq: square lattice, Ho: honeycomb lattice and Cu: cubic lattice. And for different loop-parity sectors: $\pi=\pm$ random loop-parity sector, and $\pi=+$ positive loop-parity sector.}
    \label{fig:generalized_dimension}
\end{figure}

The generalized dimension $D_q$ is defined in terms of the $\tau$-exponents as $D_q = \tau(q) / q-1$. In fig~\ref{fig:generalized_dimension} we plot the generalized dimensions as a function of $q$ for all the three cases of latices geometries and loop-parity phase space sectors. 

Furthermore, we provide some details on our calculations of $\tau(q)$-exponents defined in~\eqref{eq:tau_q}. We first stress that we use only the singular part of the spectral data for our calculations i.e. $A_{n,p}$. To better understand why one needs to remove the continuous part, it is useful to look at the relative weight of the singular part, which we can define as $\mathrm{RW} = \sum_{p,n} A_{n,p}= \sum_p B_{p}$.
Due to our normalization convention, $\mathrm{RW}\in[0,1]$.
We report the relative weights $\mathrm{RW}$ for the three lattices and different parities in the following table:

\begin{center}
    \begin{tabular}{|c|c|c|}
        \hline 
           & $\pi=\pm$ & $\pi =+$  \\
        \hline 
         Sq & $\sim 0.768$ & $\sim 0.841$ \\
         Ho & $\sim 0.071$ & $\sim 0.058$ \\
         Cu & $\sim 1.000$ & $\sim 1.000$ \\
         \hline
    \end{tabular}
\end{center}

We note that for the honeycomb lattice, most of the weight of the data is in the continuos part, while the singular multifractal part is a small but non-negligible fraction. On the other hand, the spectrum is fully singular for the cubic case. Taking the relative weights in mind, it's clear that taking the full data-set for the $\tau$-exponent calculation (at least in the honeycomb lattice) will lead to distorted results, as the non-fractal continuous part will have a significant influence on the calculations.

\begin{figure}
    \centering
    \includegraphics[width=0.75\linewidth]{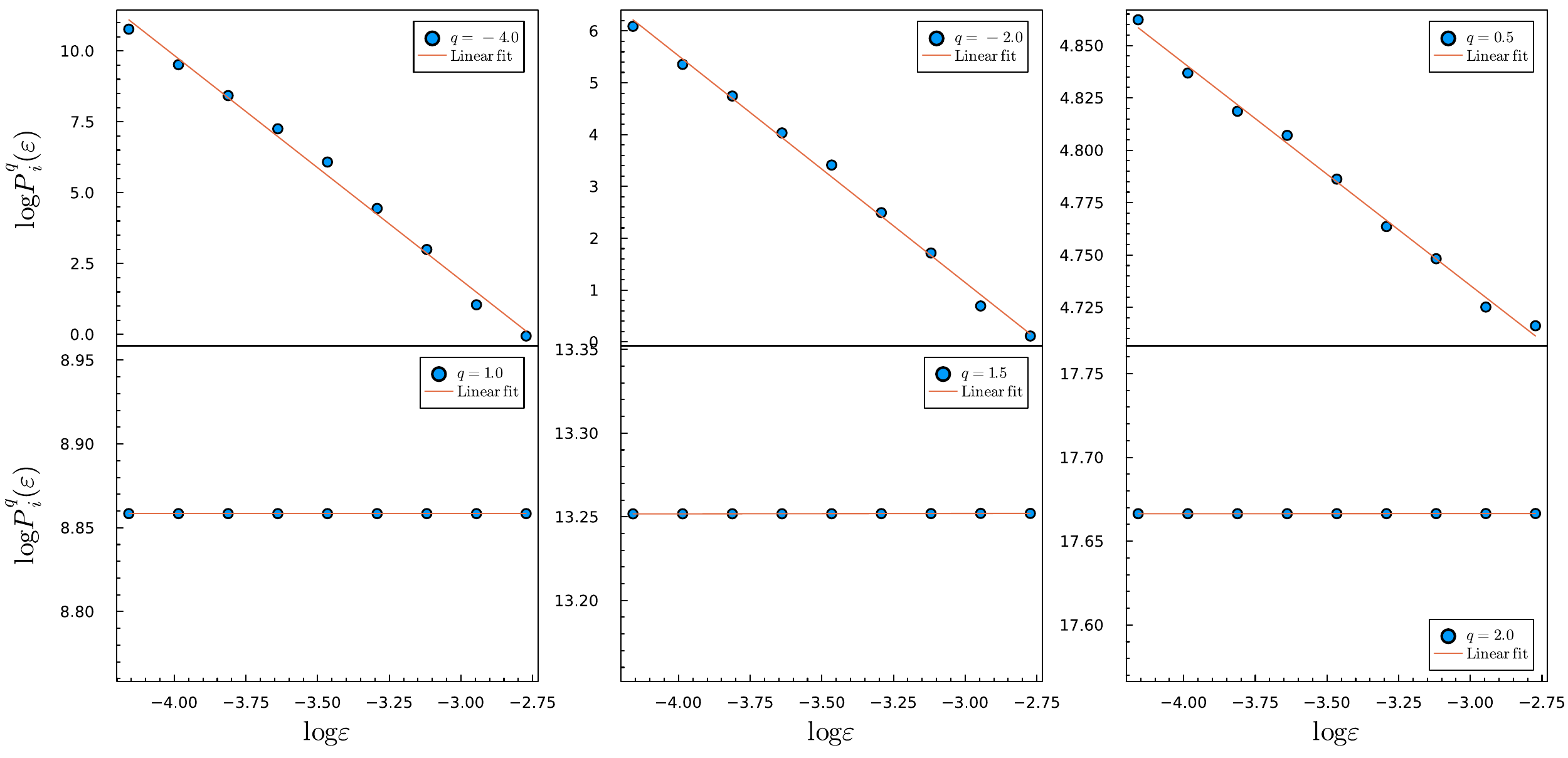}
    \caption{$\log \sum_i [P_i(\varepsilon)]^q$ vs $\log \varepsilon$ in blue for the case of honeycomb lattice for $q=-4,-2,0.5,1,1.5,2$. The orange line shows the best linear fit. 
    }
    \label{fig:tau_scaling_1}
\end{figure}

To determine, $\tau(q)$ we follow the methods outlined in~\cite{Chhabra89}, namely we take $\tau(q)$ to be the slope of the linear fit of $\log \sum_i [P_i(\varepsilon)]^q$ vs $\log \varepsilon$. Where $P_i(\varepsilon) = \int_{i\varepsilon}^{(i+1)\varepsilon} {\rm d}\omega S_{\rm sing}(\omega)$, $\varepsilon$. In fig~\ref{fig:tau_scaling_1} we demonstrate clear linear scaling for the honeycomb lattice for values of $q\in (-4,2)$.

\section{Real-time autocorrelation functions}

\begin{figure}
    \centering
    \subfloat[]{\includegraphics[width=0.75\linewidth]{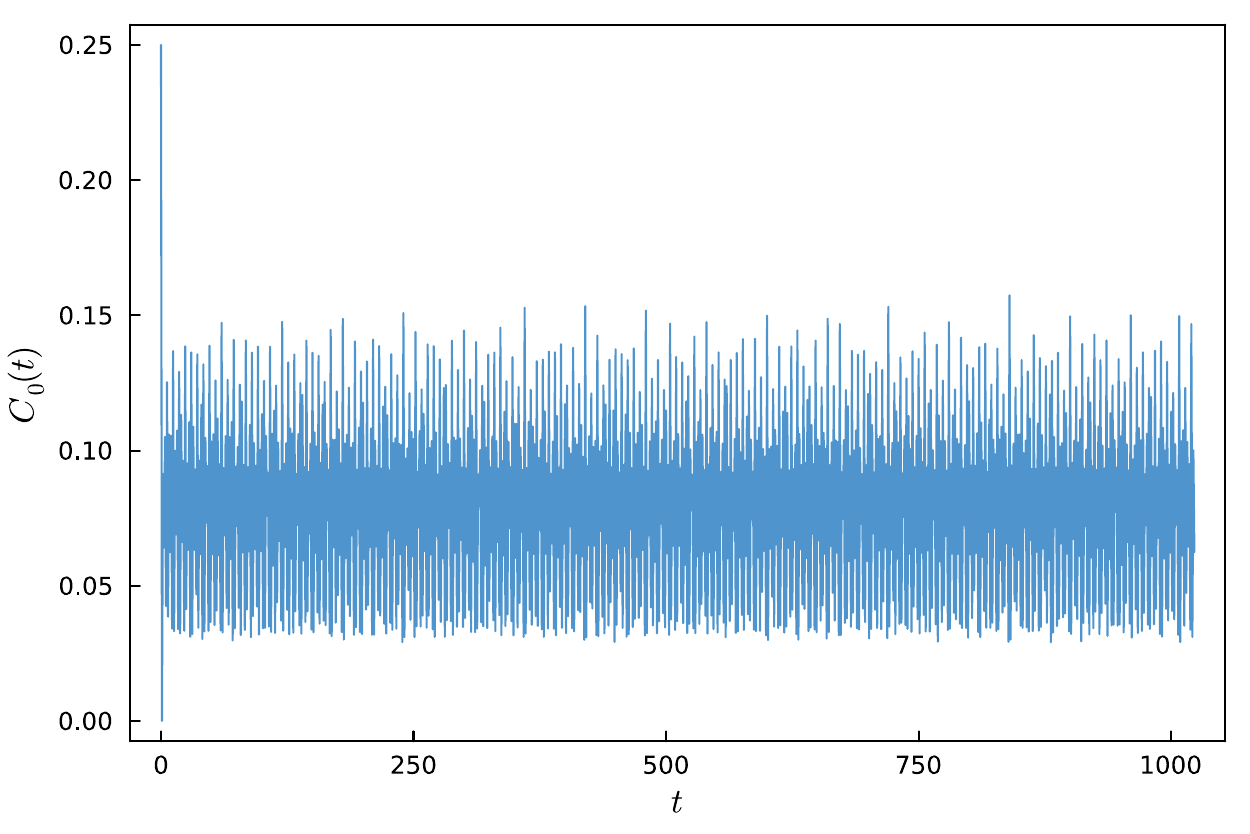}}\\
	\subfloat[]{\includegraphics[width=0.75\linewidth]{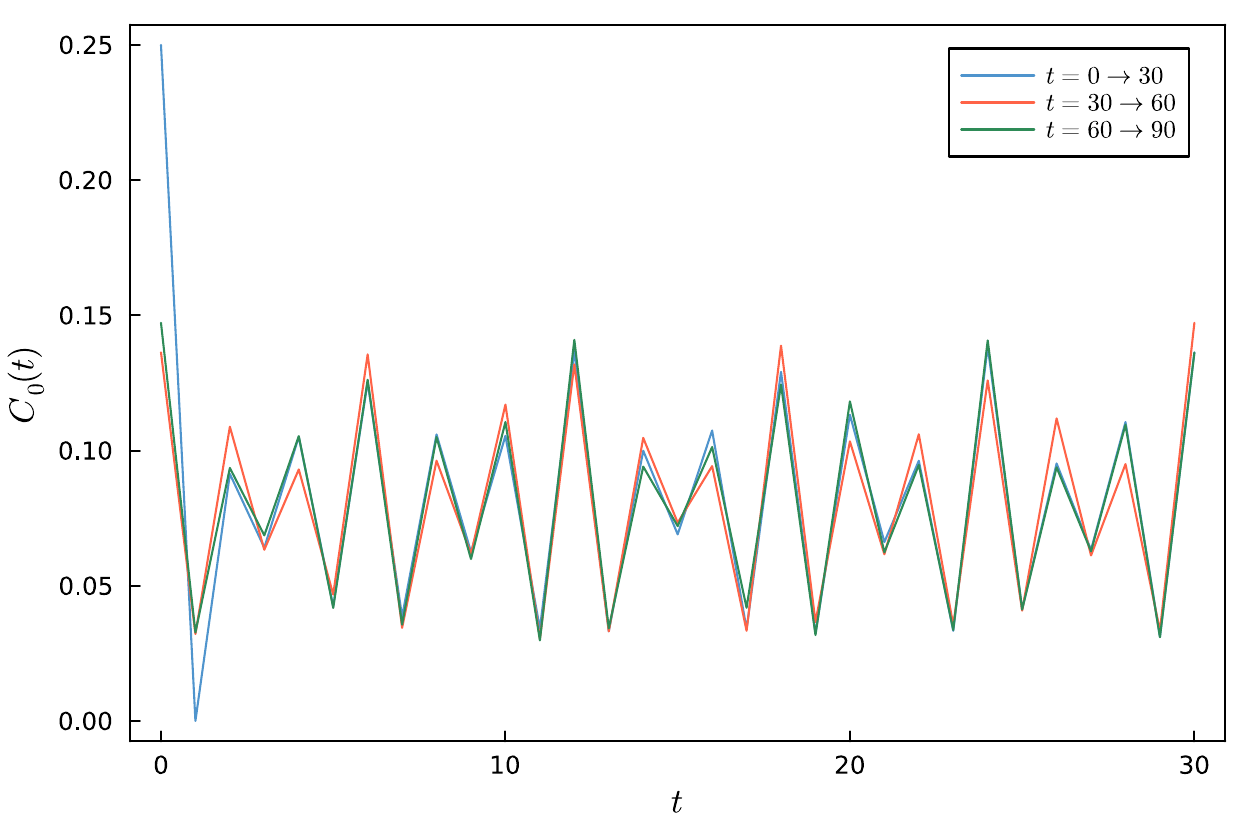}}
    \caption{Density-Density autocorrelation function for the periodic square lattice of size $64 \times 64$, with $N=2\times 64^2$ links. Panel (a) shows the autocorrelation up to time $T=1024$, and panel (b) shows near overlap of the autocorrelation functions for different time windows of fixed length $30$: $[0,30]$, $[30,60]$, $[60,90]$.}
    \label{fig:Correlation}
\end{figure}

Finally, it may be helpful to illustrate the all-periodic multifractal response by examining the real-time correlation functions. 
 In figure~\ref{fig:Correlation}(a) we plot the density autocorrelation function at $\vec{r}=0$ for a $64\times 64$ square lattice up to time $T=1024$. In figure~\ref{fig:Correlation}(b) we zoom on the section of $t\in[0,30]$ we then overlay other time sections of similar length. From the zoomed plot, one can observe the near-but-not-exact-periodicity of each segment. Furthermore, we note an approximate mirror symmetry around time-reflection at the center of time interval.

\end{document}